\documentclass[aps,prl,twocolumn]{revtex4}

\usepackage{graphicx}

\usepackage{amsmath}
\newcommand{\EqLabel}[1] { \label{#1} }

\begin{document}
 
\title{Electronic polarons and bipolarons
  in Fe-based superconductors: a pairing mechanism}

\author{Mona Berciu, Ilya Elfimov and George A. Sawatzky} 
\affiliation{Department of Physics and Astronomy, University of
British Columbia, Vancouver B.C. V6T 1Z1, Canada}

\maketitle

{\bf Superconductivity is a fascinating example of how ``more is
different''~\cite{PWA}. It is due to electrons binding into bosonic
Cooper pairs, which exhibit coherent behavior across a macroscopic
sample. Finding the mechanism responsible for this binding is one of
the more difficult tasks of condensed matter physics. For conventional
superconductors the solution was given by the BCS theory~\cite{BCS}
as being due to exchange of phonons. For the cuprate high-T$_c$
superconductors a widely-accepted explanation is still missing despite
intense effort. The recently discovered Fe-based high-T$_c$
superconductors pose now a new
challenge~\cite{Fe1,Fe2,Fe3,Fe4,Fe5}. We present here a quantum
mechanical theory for pnictides describing the influence of the large
electronic polarizability of the heavy anions. We demonstrate that its
inclusion results in electronic polarons as the low-energy
quasi-particles and also unveils a pairing mechanism for these
electronic polarons. }

\begin{figure}[t]
\includegraphics[width=0.90\columnwidth]{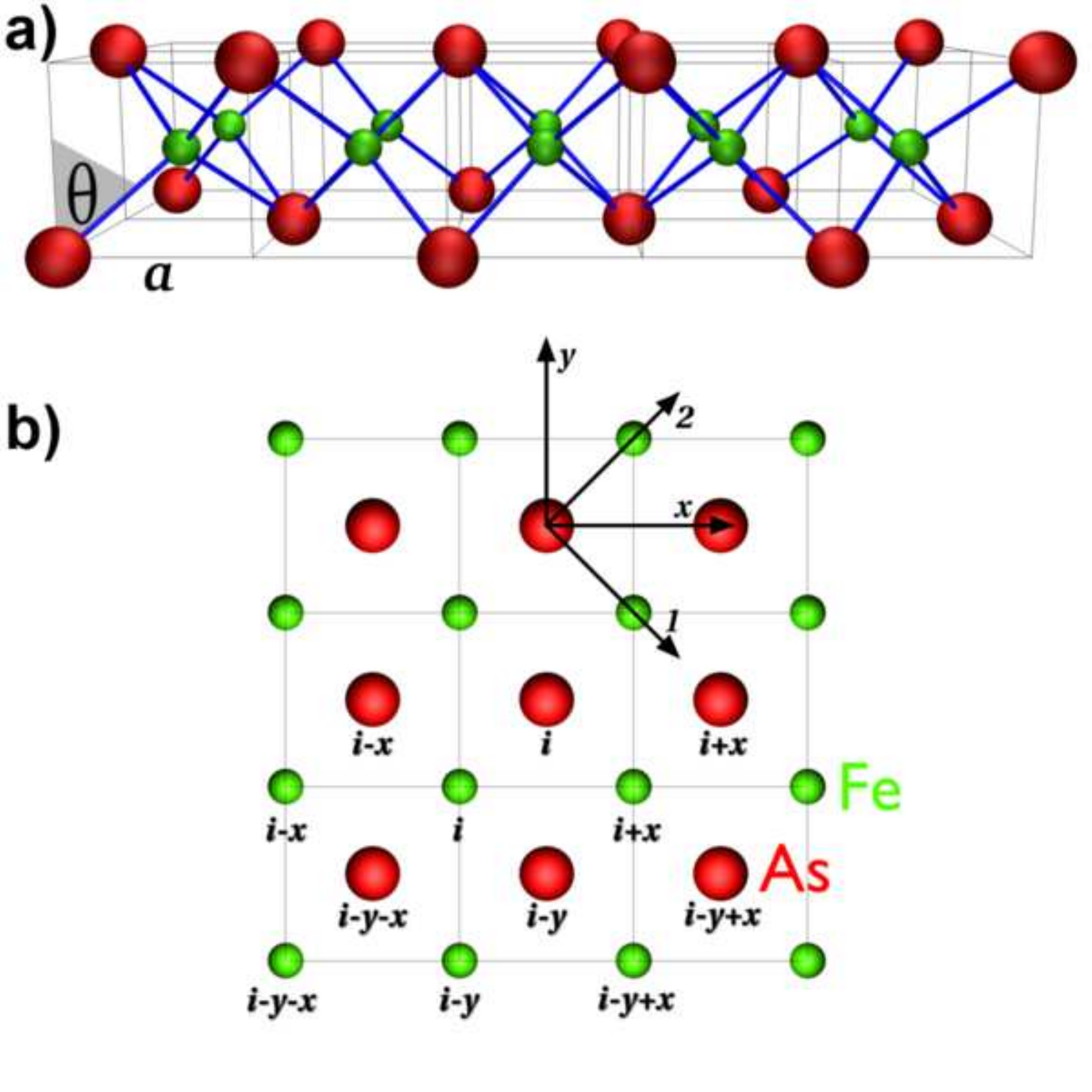}
\caption{(a) 3D sketch of the idealized FeAs layer. The lattice
  constant $a=2.8\AA$ and angle $\theta$ (see text) are indicated; (b) Top
  view of the same Fe-As layer. Several sites are 
  indexed.  The Fe and As atoms are at different ``depths''.} 
\label{fig1}
\end{figure}

At first sight, one may expect the pairing mechanism of pnictides to
be related to that of the cuprates, given the somewhat similar layered
structures.  However, there is an essential difference between the
CuO$_2$ layer where the important physics takes place in a cuprate,
and its counterpart, the FeAs layer of a pnictide: while the former is
two-dimensional (2D), with all Cu and O atoms in the same layer, the
latter is not. Instead, the layer hosting the Fe atoms (which are
arranged on a simple square lattice) is sandwiched between two layers
which share equally the As atoms, as sketched in Fig. \ref{fig1}(a).
Each Fe has 4 nearest neighbor (nn) As atoms at a distance
$R=2.4\AA$, arranged in a somewhat distorted tetrahedron, two in the
upper and two in the lower layer.

The different geometry has  important consequences. In the
cuprates, the states near the Fermi energy consist of strongly
hybridized Cu $3d$ and O $2p$ orbitals. In contrast, hardly
any hybridization appears between the Fe $3d$  and As $4p$
orbitals, with the former giving essentially all the contribution to
the low-energy states of the pnictides~\cite{G-old}. This lack of
hybridization is due not only to the lattice structure but also to
the substantial 
spread in space of  $4p$ orbitals, as compared to $2p$ orbitals. 

As a result, one might assume that the simplest Hamiltonian 
describing the low-energy physics  of pnictides  is a Hubbard
Hamiltonian for the Fe $3d$ electrons, namely:
\begin{equation}
\EqLabel{e1}
{\cal H}_{\rm Fe}= -\sum_{i,j,
  \sigma}^{} \left(t_{ij} 
c_{i,\sigma}^\dagger c_{j,\sigma} + h.c.\right)+U_H \sum_{i}^{}
\hat{n}_{i\uparrow} \hat{n}_{i\downarrow} 
\end{equation}
where $c^\dagger_{i,\sigma}$ creates an electron on the Fe site
$i$ with spin $\sigma$ and $\hat{n}_{i\sigma} =
c^\dagger_{i\sigma}c_{i\sigma}$. For simplicity, here we only count
``doping'' electrons, i.e. extra charges on top of the $3d^6$
configuration of the Fe in the undoped compound. Of course, a
multiple-band Hamiltonian can also be used 
for the proper  description of the various $3d$ orbitals, however the essential
physics we want to discuss is already captured within this simpler starting
point. The hopping integral is certainly finite for nn Fe sites, with
$t_{i, i+x}=t$. We will also consider the effects of including 2$^{nd}$
nn hopping, with $t_{i, i+x+y}=t'$ [the indexing of
various sites is indicated in Fig. \ref{fig1}(b)]. The hopping
to As sites is neglected, given the small hybridization.

Of course, Hamiltonian (\ref{e1}) is the most common starting point
for cuprates (where it does include the O 
contribution, as  $c^\dagger_{i,\sigma}$ are operators for
Zhang-Rice singlets~\cite{ZR}). In contrast, using (\ref{e1}) or a
similar starting point either in the
strong-coupling~\cite{S1,S2,S3,S4,S5,S6} or weak-coupling 
limit~\cite{W1,W2,W3,W4}  as the 
low-energy Hamiltonian for pnictides implies that the As
anions play no role in their physics.

However, the As atoms are not irrelevant. As
pointed out recently in Ref. \onlinecite{G-old}, these are big,
highly-polarizable ions which are strongly
influenced by the extra charges in their vicinity. One expects each
such charge to be  surrounded by polarized As ions,
giving rise to {\em electronic polarons}. This
results in a strong screening of the on-site Coulomb repulsion,
suggesting that these materials are not in the large $U$ limit of a
Mott-Hubbard insulator.  It also 
results in a strong nn attraction, which may be 
the key component in the pairing mechanism. The arguments of
Ref. \onlinecite{G-old} are 
based on semi-classical estimates. Here we propose a
quantum model which allows us to investigate this phenomenology away
from the linear regime, and  to also study dynamic properties of these
polarons and bound bipolarons.

The polarization of the As$^{3-}$ ions by the electric fields of the external 
charges is due to matrix elements coupling  their filled $4p$ to 
(primarily) their empty $5s$ orbitals (more details are in the
supplementary material). This is equivalent to a hole being
virtually excited from the $5s$ into the $4p$ orbitals. We therefore describe
each As by 4 operators: $s^\dagger_\sigma, p_{x,\sigma}^\dagger,
p_{y,\sigma}^\dagger, p_{z,\sigma}^\dagger$, which create a hole of
spin $\sigma$ in the respective orbital. The ground-state of an As
ion is $s_\uparrow^\dagger s_\downarrow^\dagger|0\rangle$ and the
$p$-orbitals lie at an energy $\Omega$ above it, so that the
unperturbed As ions are described by:
\begin{equation}
\EqLabel{e2}
{\cal H}_{\rm As} = \Omega \sum_{i,\lambda,\sigma}^{}
p^\dagger_{i,\lambda,\sigma} p_{i,\lambda,\sigma},  
\end{equation}
where $i$ is the location of the As ion [see Fig. \ref{fig1}(b)]
$\lambda=x,y,z$ and $\sigma$ is the spin of the hole. Hopping between
 As ions is very small, given the large nn distance between
them of about $4\AA$, and we ignore it. Indeed, LDA
calculations find narrow $4p$ bands~\cite{G-old}, justifying this
atomistic description of the As ions.

We make two approximations regarding the interactions: (i) only
As atoms nn to an Fe hosting a charge are polarized by its
electric field, and (ii) we ignore dipole-dipole interactions between
As clouds. (Estimates of the
corresponding corrections are given in the supplementary material).  
Our interaction Hamiltonian is, then:
\begin{eqnarray}
\nonumber  {\cal H}_{\rm int} &&= g \sum_{i, \sigma}^{} \hat{n}_i
\left[s_{i,\sigma}^\dagger\left(-\sin \theta p_{i,2,\sigma} + \cos
\theta p_{i,3,\sigma}\right)\right.\\
\nonumber  &&+
s_{i-y,\sigma}^\dagger\left(-\sin \theta p_{i-y,1,\sigma} - \cos
\theta p_{i-y,3,\sigma}\right)   \\
\nonumber  && + s_{i-x-y,\sigma}^\dagger\left(\sin \theta
p_{i-x-y,2,\sigma} + \cos \theta p_{i-x-y,3,\sigma}\right)\\
\EqLabel{e3} &&\left.+ s_{i-x,\sigma}^\dagger\left(\sin \theta
p_{i-x,1,\sigma} - \cos \theta p_{i-x,3,\sigma}\right)+ h.c.\right]
\end{eqnarray}
where the sum is over all unit cells $i$ in the lattice. This
interaction 
describes the $5s\leftrightarrow 4p$ transitions resulting in the polarization
of the As  along the appropriate Fe-As direction (see
Fig. \ref{fig1}) if doping charges, 
counted  by $\hat{n}_i = \hat{n}_{i\uparrow} + \hat{n}_{i\downarrow}$,
are on any  Fe$_i$ site. 

Our model Hamiltonian ${\cal H} = {\cal H}_{\rm Fe} + {\cal H}_{\rm
  As} + {\cal H}_{\rm int}$ is characterized by 5 energy scales. The
hopping integral $t=0.25$eV and the $5s-4p$ energy difference
$\Omega=6$eV are estimated from LDA results (for details on
determining all the parameters, see the supplementary material). We
use either $t'=0$ or $t'=-t/2$, the latter being the appropriate value
for hopping between $d_{xy}, d_{xz}, d_{yz}$ orbitals which contribute
most near the Fermi level. The interaction energy $g=2.5$eV is
extracted from the As$^{3-}$ polarizability $\alpha_p=10\AA^3$, and the
Hubbard onsite repulsion $U_H$ is left as a free parameter. We present
results for a wide range of these parameters, proving that
our results are only weakly sensitive to  their precise values.

\begin{figure}[t]
\includegraphics[width=0.90\columnwidth]{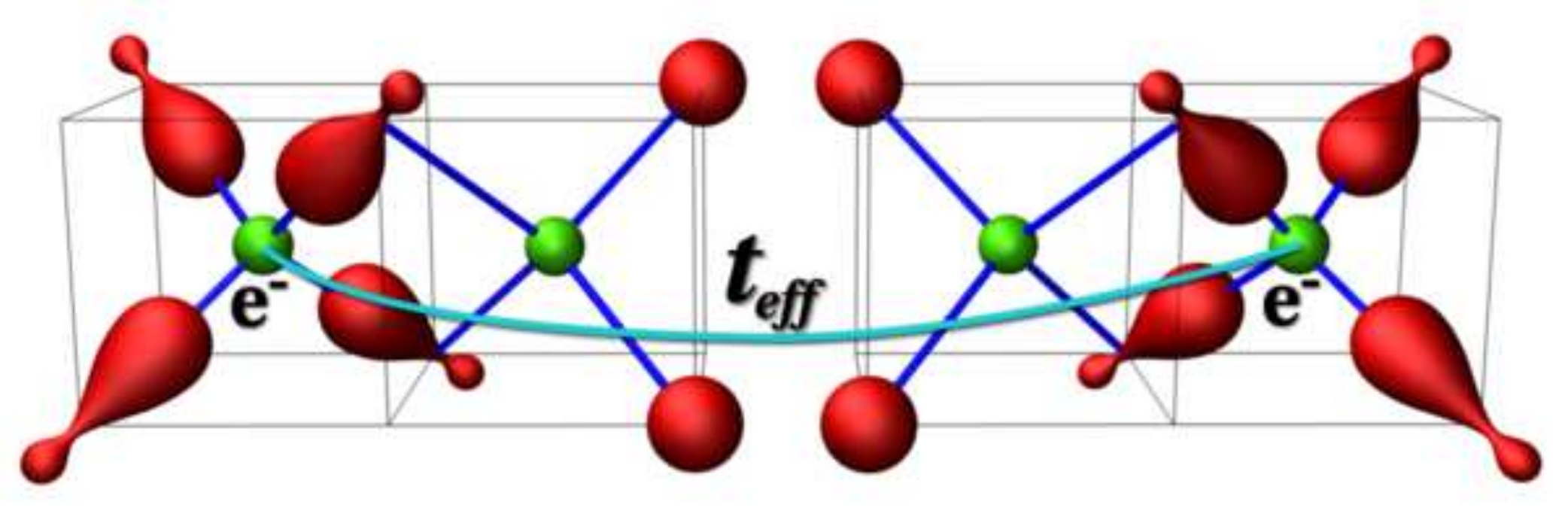}
\caption{Sketch of a single electronic polaron and its nn hopping. The 
electric  field created by the extra electron residing on the Fe excites
holes of nn As into the $4p$ orbital pointing towards the Fe (more
distant As atoms stay in the $5s$ ground state). The
hopping integral  is renormalized by the overlap of the
polarization clouds in the 
initial and final states. }
\label{fig2}
\end{figure}

We begin with a study of the low-energy spectrum of the electronic
  polaron [sketched 
  in Fig. \ref{fig2}] created when there is one doping electron in the
  system. We use  
first order perturbation theory in the hopping, which is
the smallest energy  (a quantitative justification and full
details of the calculations are provided in the supplementary material). The
  polaron eigenenergies are:
\begin{equation}
\EqLabel{e4}
E_{P}(\vec{k}) = 4(\Omega -\sqrt{\Omega^2+4g^2}) + \epsilon_{\rm eff}(\vec{k}).
\end{equation}
The first term is the interaction energy between the
electron and the induced As dipole moments. For $g\ll \Omega$ it equals $-4
[\alpha_p E^2/2]\approx $-8.2eV, $E$ being the electric field at the
As sites. This linear expression was used in
Ref. \onlinecite{G-old}. Since in fact $g/\Omega \approx 0.4$, non-linear
effects become important and reduce it slightly to $\approx-7.1$eV.
$\epsilon_{\rm eff}(\vec{k})=- 2 t_{\rm eff} \left[ \cos (k_x 
  a) + \cos 
(k_ya)\right] -4t'_{\rm eff}\cos (k_x a) \cos (k_y a)$ is a dispersion
identical to that of a 
free particle, 
but with 
renormalized hopping integrals due to the overlap of the As clouds
as the polaron moves between different sites (see
Fig. \ref{fig2}). Expressions for $t_{\rm eff}$ and $t'_{\rm eff}$ are
given in Eqs. (23), (24) in the supplementary material. We plot their
values in Fig.~\ref{fig3}, for a wide range of $\Omega$ and $\alpha_p$
values.

\begin{figure}[t]
\includegraphics[width=0.80\columnwidth]{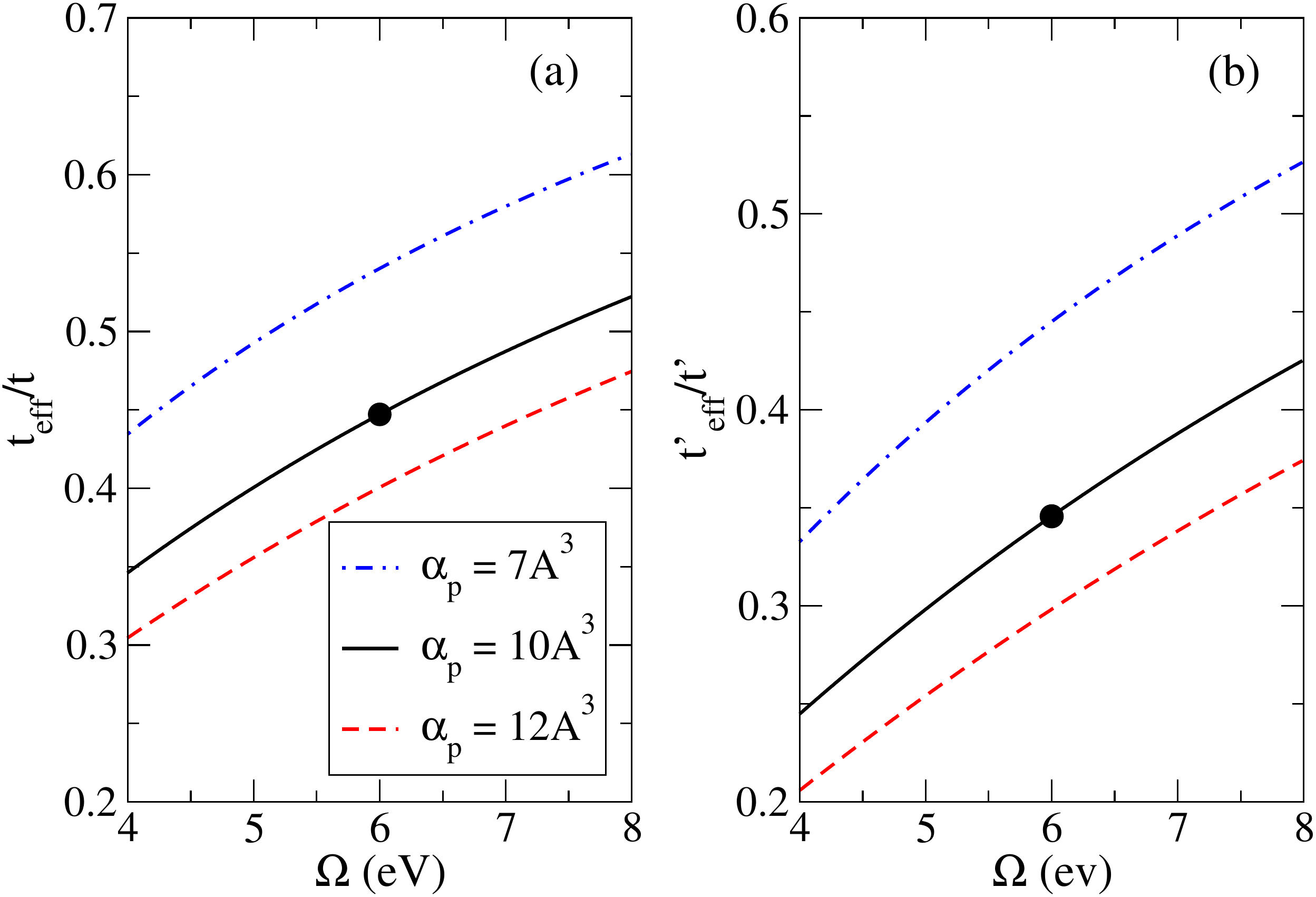}
\caption{(a) $t_{\rm eff}/t$  and (b) $t'_{\rm eff}/t'$ 
  vs. $\Omega$, for a polarizability $\alpha_p=7, 10$ and
  $12\AA^3$. The dots show the values used here.} 
\label{fig3}
\end{figure}

For a fixed $\alpha_p$, a smaller $\Omega$ implies a larger $g/\Omega$
ratio [Eq. (14) in supplementary material], {\em i.e.} a stronger effective
coupling. This explains the smaller $t_{\rm eff}$, $t'_{\rm eff}$
as $\Omega$ decreases to be typical polaronic
physics: stronger interactions lead to heavier
polarons~\cite{polarons}. However, note that these ratios are 
not exponential in the coupling, like for normal lattice polarons, but rather
weakly dependent on the various parameters. This is due to non-linear
effects, namely the existence of a maximum allowed value of the induced
dipole moment.  Both $t_{\rm eff}/t$ and $t'_{\rm eff}/t'$ are around
50\% for a wide range
of values centered on our parameters, suggesting a light polaron with
an effective mass roughly twice as big as the free band carrier.

Such renormalization of the band-structure compared to that predicted
by LDA should be detected by ARPES. Interestingly, a recent
measurement~\cite{ARPES} showed reasonable fits to LDA upon rescaling
the bandwidth by a
factor of 2.2, similar to our typical values (the material studied has
P instead of As, and these smaller anions have a smaller polarizability
$\alpha_p \approx 7\AA^3$). To first order in perturbation theory, we
also predict a quasiparticle weight which is independent of $\vec{k}$,
with $Z=0.38$ for our parameters. The remaining spectral weight
should be observed at much high energies typical of the excited As
polarization clouds, of the order ${1\over 2} \left[ \Omega+
\sqrt{\Omega^2 + 4g^2}\right] = 5eV$ and higher (more details are in
the supplementary material).

We now study a system with 2 doping electrons, also within first
order perturbation  in the hopping. We focus on singlet solutions
(eigenstates are either singlets or
triplets of the two electrons). Also, we first analyze results for
$t'=0$, then investigate the role  of a finite $t'$.

If there were no interactions between the two polarons, then for a
total momentum $\vec{k}$ one would get a continuum of eigenstates with
energies $E_P(\vec{k}-\vec{q}) + E_P(\vec{q})$ for all $\vec{q}$ in
the Brillouin zone, with the ground-state at
$\vec{k}=\vec{q}=0$. However, there are short range interactions
between the polarons. If the charges are 2$^{nd}$ nn or closer to each
other, some As ions are simultaneously nn to both charges and acquire
a dipole moment different from that of As ions which interact with a
single charge. This leads to different energies $U_0$, $U_1$ and $U_2$
for configurations with the two charges being, respectively,
on-site, nn and 2$^{nd}$ nn to each other.  We sketch below the
derivation of $U_0$ and $U_1$. The effective hopping 
between these configurations is also different from $t_{\rm eff}$,
because of the overlap with these differently polarized clouds.
These
interaction energies and some of the effective hoppings are summarized
in Fig.~\ref{fig4} (expressions and plots for all these quantities
are provided in the 
supplementary material).

\begin{figure}[t]
\includegraphics[width=\columnwidth]{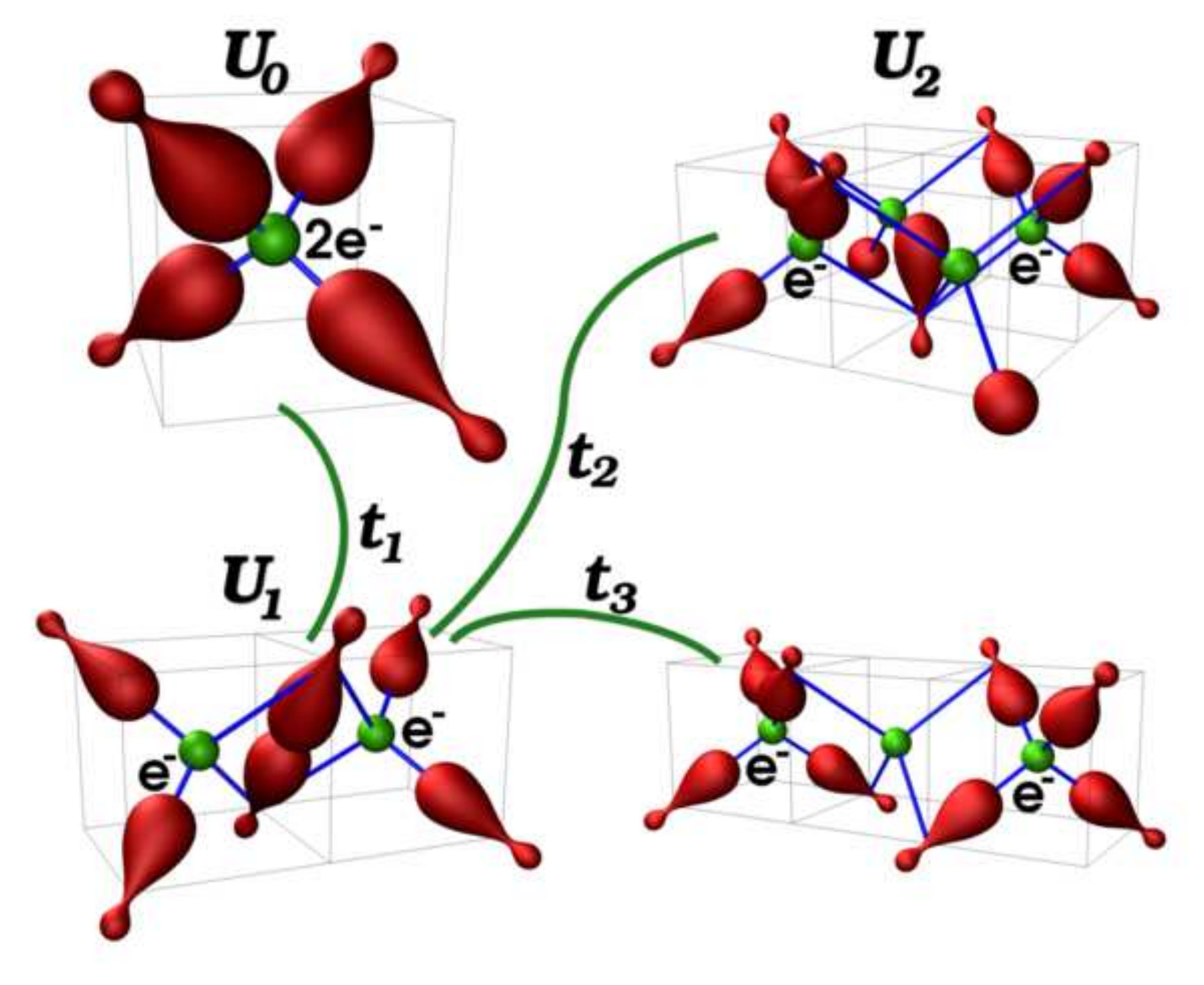}
\caption{Sketches of on-site, $1^{st}$, $2^{nd}$ and $3^{rd}$ nn
  bipolarons. The first three configurations have
  interaction energies 
  $U_0, U_1$ and $ U_2$, respectively. Several of the special effective
  hopping integrals 
  are also indicated.} 
\label{fig4}
\end{figure}

\begin{figure}[t]
\includegraphics[width=\columnwidth]{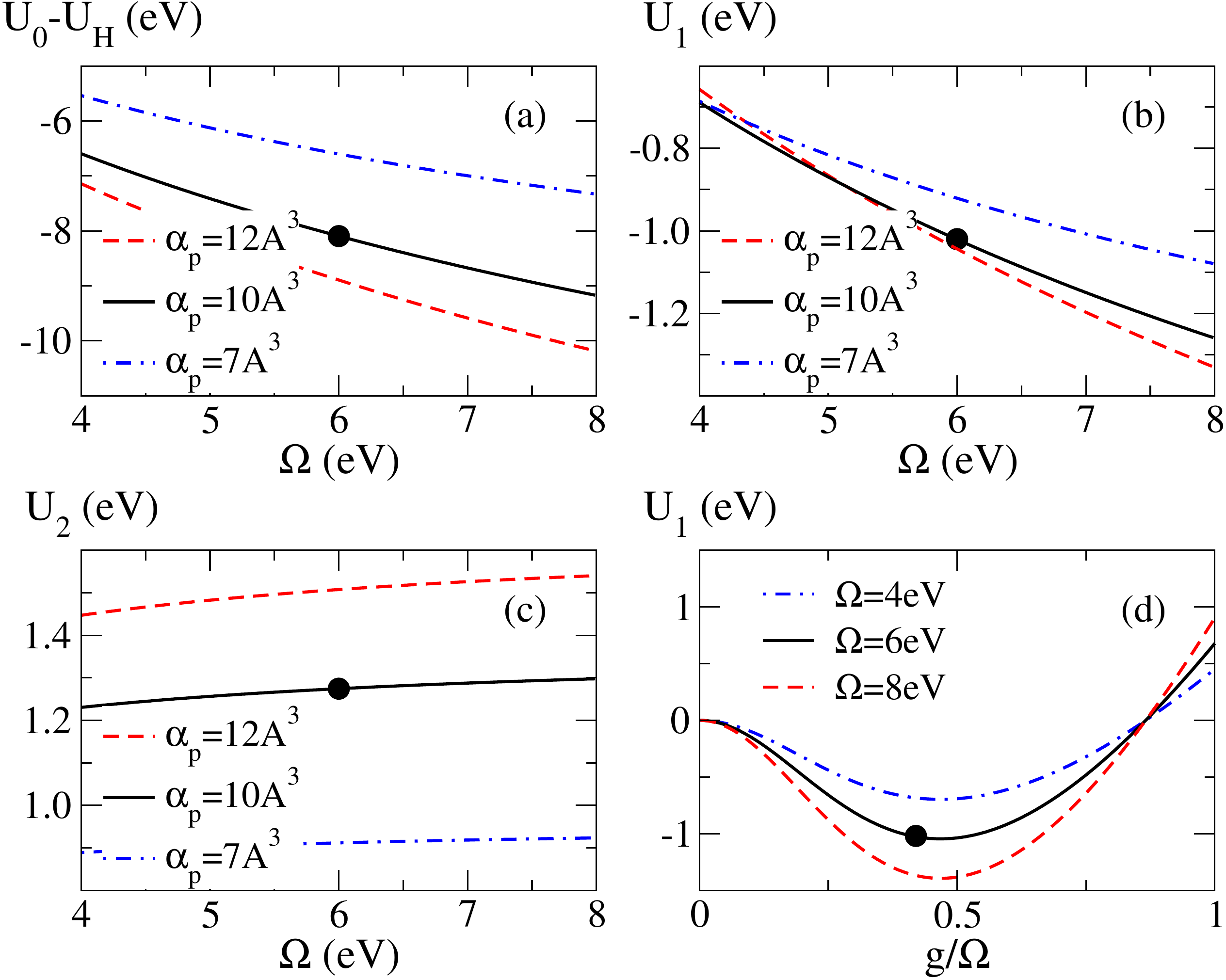}
\caption{(a) Renormalization of on-site interaction, $U_0-U_H$; (b) nn
energy $U_1$ and (c) $2^{nd}$ nn energy $U_2$ vs. $\Omega$ for various
polarizabilities. (d) $U_1$ vs. $g/\Omega$  when
$\Omega=4,6,8$eV. The dots show our typical values.} 
\label{fig5}
\end{figure}

An on-site bipolaron also has 4 polarization clouds, however 
effectively $g\rightarrow 2g$ in Eq. (\ref{e3}), as the on-site charge
is doubled, so its electrostatic energy can be immediately obtained from
Eq. (\ref{e4}) to be $E_{BP,0} = U_H + 4(\Omega
-\sqrt{\Omega^2+16g^2})$. The energy of two static distant polarons is
$E_{BP,\infty}=8(\Omega -\sqrt{\Omega^2+16g^2})$ , therefore $U_0 =
E_{BP,0} - E_{BP,\infty} = U_{\rm H}- 4\left[\sqrt{\Omega^2 +16g^2} +
\Omega - 2 \sqrt{\Omega^2 +4g^2}\right]< U_H$ always. This is obvious
in the linear regime: because the electric field is twice as large,
the energy of the bipolaron is 4 times  that of a polaron,
{\em i.e.} twice as large as of two free polarons. This leads to a
significant screening of $U_H$, of several eV, as shown in
Fig. \ref{fig5}(a) (these values are smaller than in
Ref. \onlinecite{G-old} because of non-linear effects). Such a
strong on-site attraction is well-known to arise for polarons in
general~\cite{Alex}. 

A nn bipolaron has 6 polarization clouds 
(see Fig.~\ref{fig4}), the two central ones being larger than the
regular polaron clouds. In the
linear regime, the energy of each central cloud is
$-\alpha_p (\vec{E}_1+\vec{E}_2)^2/2$, so the energy difference with
respect to two regular clouds is $-\alpha_p
\vec{E}_1\cdot\vec{E}_2<0$ if the Fe-As-Fe angle is less than 90$^o$,
as is  here. This means an attractive nn interaction, as indeed
seen for typical parameters in Fig. \ref{fig5}(b). However, because of
non-linear effects, this interaction becomes repulsive at strong
coupling, as shown in Fig. \ref{fig5}(d). Interestingly, this figure
reveals that these materials are close to optimal, with $U_1$ near its
minimum, for typical values of the parameters (see also Fig. 3 in
the supplementary material).

This nn attraction {\em is not} typical polaron physics. On a 2D
lattice like in cuprates, the interaction between electronic polarons
is {\em repulsive} because 
the Cu-O-Cu angle is 180$^o$ (the same holds true for  polarons
coupled to a breathing-mode phonon of the O atoms). For a Holstein
bipolaron, there 
is a weak nn attraction, but it is due to spin exchange and is ${\cal
  O}(t^2)$, not ${\cal O}(t^0)$ like here~\cite{Alex}. Its non-2D 
geometry is the essential ingredient in bringing about this strong nn
attraction, for the Fe-based superconductors. This mechanism is
absent in cuprates.

\begin{figure}[b]
\includegraphics[width=0.8\columnwidth]{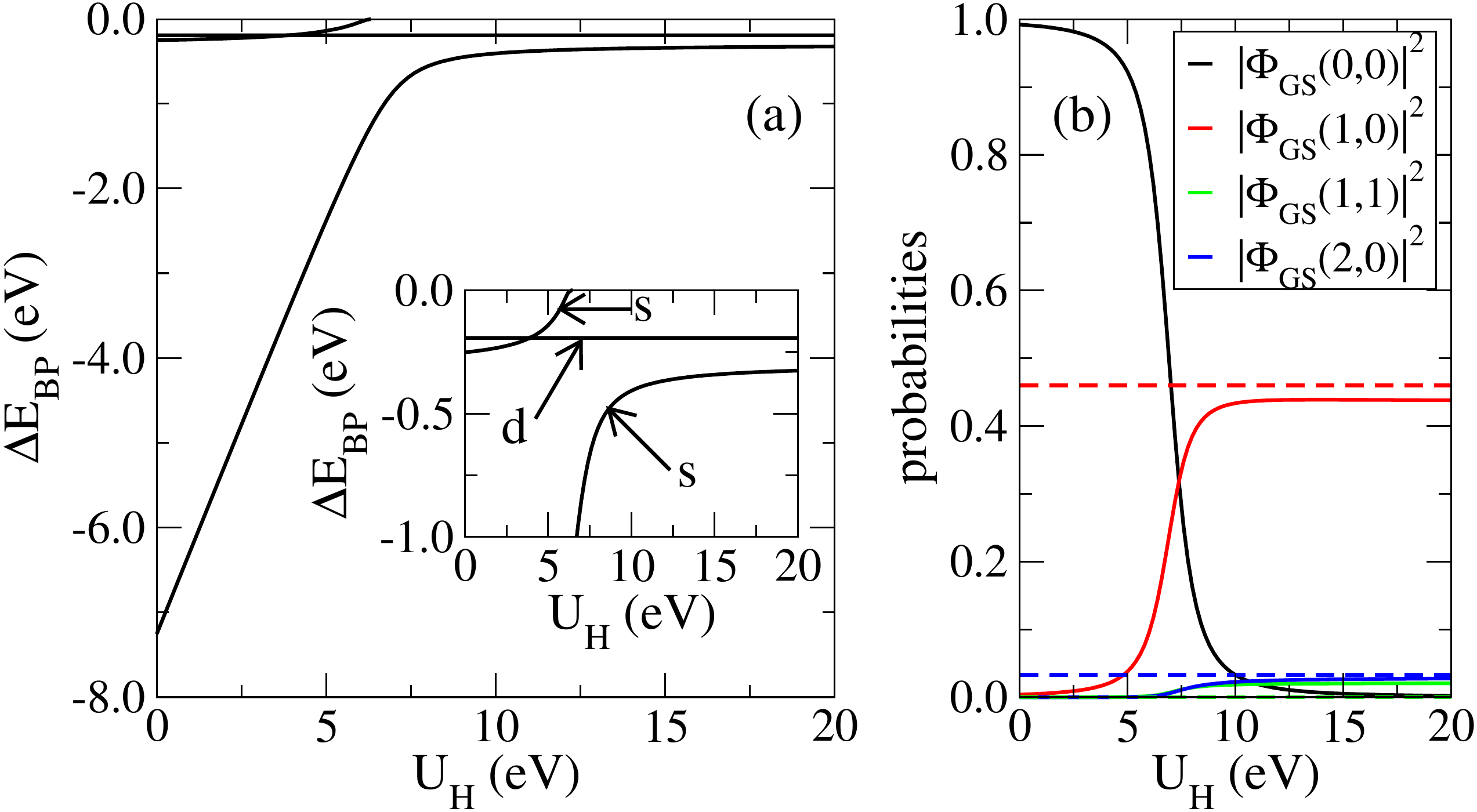}
\caption{(a) Eigenstates below the two-polaron continuum, $\Delta
  E_{BP}= E_{BP}(0)+8t_{\rm eff}$ vs. $U_H$. The symmetry of the three bound
  eigenstates is labeled in the inset. (b) Probability for on-site, $1^{st}$,
  $2^{nd}$ and $3^{rd}$ nn separation in the ground state,
  vs. $U_H$. The dashed lines show the same quantities for the
  $d$-state. We use
  $t'=0, \alpha_p=10\AA^3$, $\Omega=6eV$ (similar results are
  found for all $\alpha_p=7-12\AA^3, \Omega=4-8$ eV).
} 
\label{fig6}
\end{figure}

First order perturbation in hopping mixes  these
static configurations. Following the calculation described in the
supplementary material, we obtain the eigenergies $E_{BP}(\vec{k})$
for a total momentum $\vec{k}$. In
Fig. \ref{fig6}(a) we plot $\vec{k}=0$ eigenenergies for all bound
bipolaron states,  {\em i.e.} which have energies
below the two-polaron continuum starting at $-8t_{\rm eff}$. We find 3 such
states. The GS first shows linear dependence on $U_H$ and then
flattens out. Its wavefunction  has
$s$-type symmetry (unchanged sign upon $90^o$ rotation). There is a second 
$s$-state at small $U_H$, which then joins the
continuum. The 3$^{rd}$ bound bipolaron state has an energy 
independent of $U_H$, and is $d$-type (wavefunction changes
sign upon  $90^o$ rotation). 

The nature of these bound states is revealed in
Fig. \ref{fig6}(b). For low $U_H$ values, the GS primarily consists of
an on-site bipolaron, with hardly any contribution from nn or more
distant configuration. This explains why its energy here scales with
$U_H$ (more precisely with $U_0$). When $U_0\approx 0$ there is a fast
crossover to a state dominated by the nn bipolaron configurations (a
combined total of 90\% probability). This is expected, since $U_1 <0$
irrespective of $U_H$, favoring such a pair when $U_0$ becomes
repulsive. The onsite contribution is now exponentially small. This
explains the weak $U_H$ dependence here, as coming from virtual
hopping to the on-site configuration. The dashed lines show the
contributions for the $d$-type state. As expected, it is dominated by
nn bipolarons. It has a zero on-site probability, consistent with its
symmetry and explaining the lack of dependence on $U_H$. The $2^{nd}$
nn contribution is also zero ($U_2 >0$ as well). There are small
$3^{rd}$ and more distant bipolaron contributions.

\begin{figure}[t]
\includegraphics[width=0.9\columnwidth]{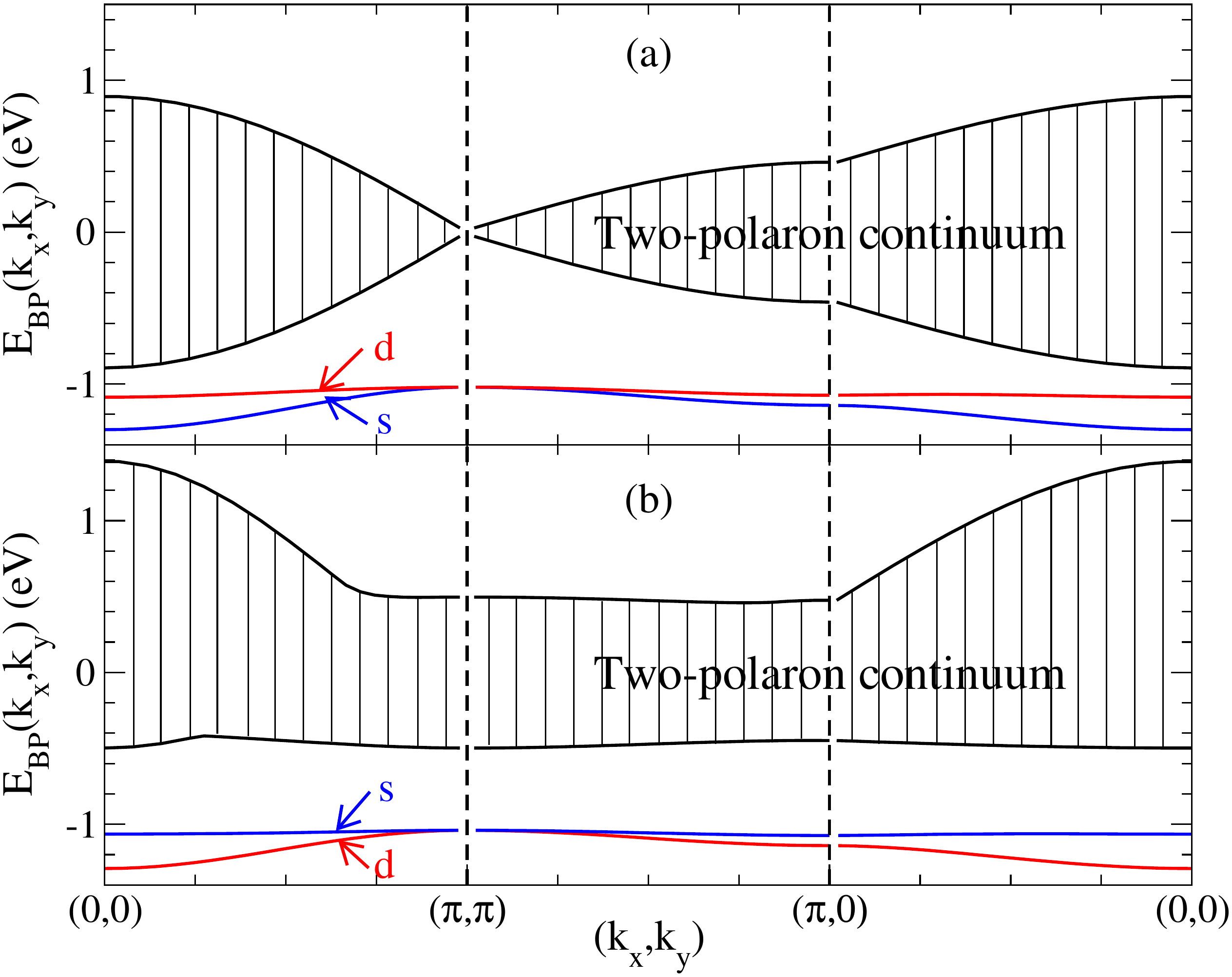}
\caption{Dispersion of the two bound bipolaron states along
  high-symmetry axes in the Brillouin zone, for (a) $t'=0$ and (b)
  $t'=-t/2$. The two-polaron continuum is also shown. Parameters are
  $U_H=10$ eV, $\alpha_p=10\AA^3, \Omega=6$ eV (similar results are
  found for all $\alpha_p=7-12\AA^3, \Omega=4-8$ eV). The symmetry of
  the ground state changes from $s$ to $d$ if $t'\ne 0$. 
} 
\label{fig7}
\end{figure}

\begin{figure}[t]
\includegraphics[width=0.8\columnwidth]{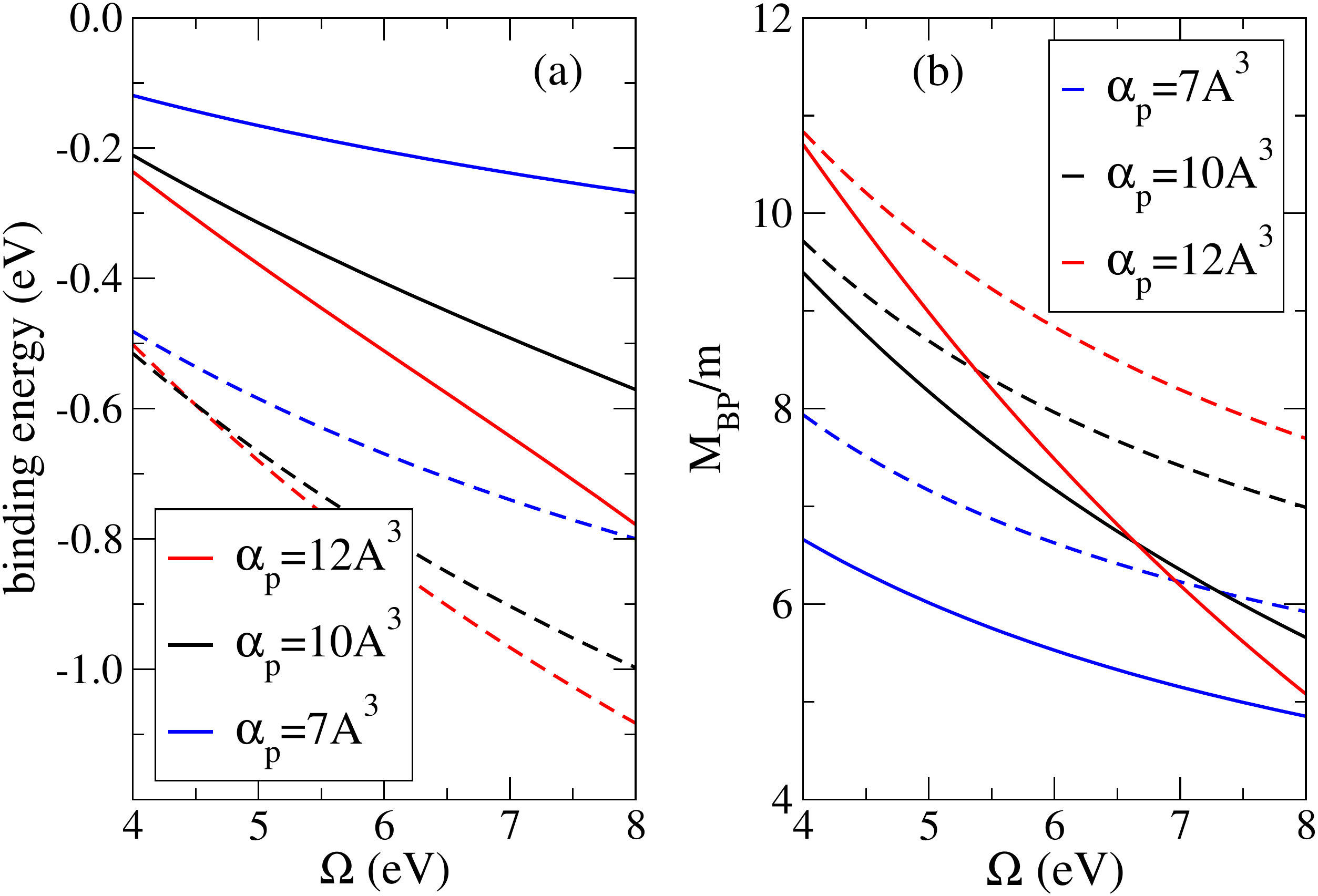}
\caption{Ground-state bipolaron (a) binding energy, and (b) effective
  mass in units of the free carrier mass vs. $\Omega$, 
  for various polarizabilities. The full lines correspond to $t'=0$,
  dashed lines to $t'=-t/2$. Here $U_H=10$eV.
} 
\label{fig8}
\end{figure}

The unscreened Hubbard repulsion is very large, $U_H\sim 10-20$
eV. For these values there is hardly any dependence on $U_H$, so its
precise value is of little importance. We use $U_H=10$ eV from
now, and ask how mobile are these bound, predominantly nn, bipolaron
pairs. Their dispersion in the Brillouin zone in shown in 
Fig. \ref{fig7}(a), where we also show the two-polaron continuum
(eigenstates above the continuum are not shown). The
$s$-pair has a bandwidth $E_{GS}(\pi,\pi)-E_{GS}(0,0) \approx 0.3$
eV, implying a bipolaron mass  about 7 times that of a free carrier
mass (the bandwidth of a free electron is $8t=2$eV). This is not a huge
enhancement, since it means that the bipolaron is about 3.5 times
heavier than a single polaron. In Fig. \ref{fig8}(b) we plot the
bipolaron mass for 
various $\alpha_p$ and $\Omega$ values, showing only limited variation
over a wide parameter range. The higher energy $d$-pair is
 heavier, with a much narrower bandwidth.

If we include 2$^{nd}$ nn hopping $t'=-t/2$, two effects are
apparent. First, the two-polaron continuum is pushed to higher
energies, effectively increasing the binding energies of both bound
bipolaron
states. (The binding energies for the GS bipolaron are shown in
Fig. \ref{fig8}(a) for various parameters).  
Second, the $d$-state becomes the ground-state. This is
not surprising, since the 2$^{nd}$ nn hopping links directly the two
nn bipolaron 
configurations which give
the bulk contribution to these eigenstates. Since $t'<0$, this mixing
raises the energy of an $s$ state and lowers that of a $d$-state. Thus, if
the effective $t'$ between the two nn bipolaron configurations is
large enough, the $d$ state has to become the ground-state.

Let us now comment briefly on the triplet eigenstates. Since there is
no onsite triplet bipolaron configuration, their energies are
essentially identical to the energies of the singlet eigenstates in
the limit $U_H\rightarrow \infty$. This implies that at this level of
approximation and for large enough $U_H$, singlet and triplet bound
bipolarons are almost degenerate. However, it is well known that 
second-order perturbation theory produces an exchange interaction
which strongly favors the singlet eigenstates (this is the interaction
responsible for binding the S1 Holstein bipolaron, see for example
Ref. \onlinecite{Alex}). It follows that in reality, the ground-state
must be a 
singlet.

The binding energies shown in Fig. \ref{fig8}(a) are substantial, even
without inclusion of this singlet exchange energy. As discussed in the
supplementary material, relaxation of our approximations (that only As
ions nn to a charge are polarized and that dipole-dipole interactions
are ignored) further enhances $U_1$, and therefore these binding
energies, to several 
eV. Yet more enhancement is expected if we include even higher orbitals
than $5s$ when describing the As ion polarization. All this suggests
the appearance of pre-formed pairs well above room 
temperature. On the other hand, in our model Hamiltonian we have
ignored a nn repulsion energy which comes from the bare Coulomb interaction
itself (in reality, this nn Coulomb repulsion will be reduced by other
screening mechanisms, such as the bond polarizabilities involving the
As $4p$ and Fe $3s$ and $4p$ states). This nn repulsion will decrease
$U_1$ substantially, and may even change its sign, making nn bipolaron
pairs unstable. It is difficult to obtain accurate quantitative
estimates of all these terms, to find out whether bound bipolaron
states exist and what are their binding energies.

It is important to point out that the presence of this As-mediated nn
 attractive interaction $U_1$ may be essential even if preformed
 bipolaron pairs do not exist. This would put these materials in a
 BCS-like framework, with the phonon glue replaced by a virtual
 excitonic glue. Because $U_1$ is so large one does not really need
 much of a retardation effect to overscreen the bare Coulomb
 repulsion. In favor of this scenario is experimental data indicating
 higher T$_c$ in samples with shorter Fe-As distances and smaller
 Fe-As-Fe angles, which is precisely what increases the value of $U_1$
 (more speculation on this is presented in the supplementary
 material).

However we do not want to rule out the presence of preformed pairs of
the kind discussed above. There is ample evidence that singlet pairs
could exist to quite high temperatures in these systems. For example,
the magnetic susceptibility~\cite{Dresden} shows the same strong
increase with temperature above the spin density wave (SDW) or
superconducting transition temperature. This is very difficult to
explain if we start from a free single-particle-like picture, but easy
to understand if we assume the presence of singlet preformed pairs
well above T$_c$, with a binding energy of the order of 100meV or
more. The magnetic susceptibility of such a system would increase with
increasing $T$ with an activated kind of behavior. NMR Knight shift
data also displays this kind of behavior~\cite{Dresden1}.

We therefore suggest to take the scenario of preformed singlet
bipolarons seriously. We note that a superconducting state would then
behave more like a Bose Einstein condensate. At first glance one
might think that its T$_c$ would have to be very low, however we note
that these bipolarons are very light, with a mass which is about 3-4
times the single polaron mass, and thus very high condensation
temperatures could be expected.  The other rather interesting aspect
to consider is that perhaps this scenario might also explain the low
amplitude SDW observed at low dopings, as being a
different kind of condensate of singlet bipolarons due to rather
strong exchange interactions. The low amplitude would be a result of
the pair-wise singlet formation tendency competing with the pair-pair
exchange interactions which would favor the SDW. We propose to study
these issues next.

\acknowledgments
Acknowledgments: We  thank J. Zaanen, J. van den Brink, C. Varma,
D. Bonn and A. Damascelli  for many
stimulating discussions and insightful opinions. This work was
supported by NSERC, by CIfAR Nanoelectronics  and Quantum Materials,
and by 
the Killam Trust 
and the Alfred P. Sloan Foundation (M.B.).

Competing interests statement: 
The authors declare no competing financial interests.

\end{document}


\title{Supplementary material}
\begin{abstract}
\noindent for the article {\em  ``Electronic polarons and bipolarons
  in Fe-based superconductors: a pairing mechanism''} by
Mona Berciu, Ilya Elfimov and George A. Sawatzky.
\end{abstract}

\maketitle

\tableofcontents

\newpage

\section{The model: lattice and Hamiltonian}

\begin{figure}[b]
\includegraphics[width=0.80\columnwidth]{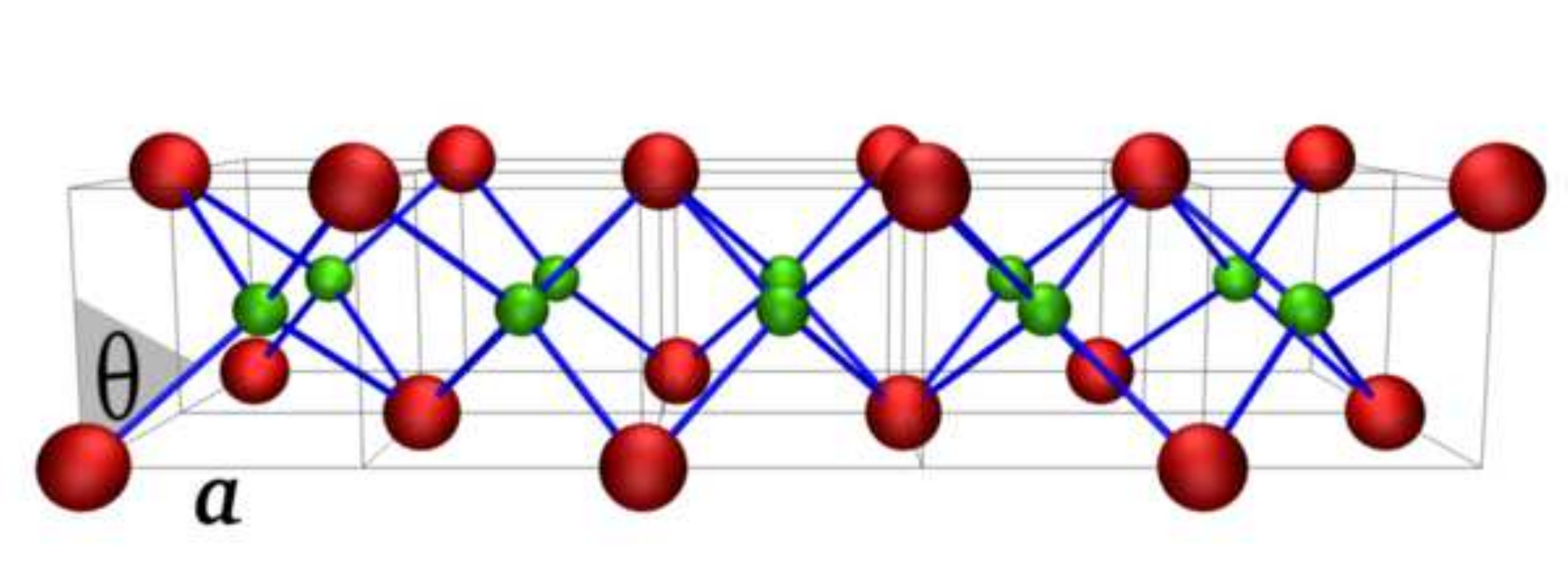}
\caption{3D sketch of the FeAs layer. The lattice constant is $a$ and
  the angle $\theta$ between the Fe-As direction and the $z$-axis is
  indicated. } 
\label{fig1}
\end{figure}

The idealized FeAs layer of interest is sketched in Fig. \ref{fig1},
which shows 
the cubic lattice (lattice constant $a=2.8\AA$) which has the Fe  in
the center of the cubes, on a simple square lattice. The As  are
divided equally between the top and bottom layers, as indicated. Each
Fe  has 4 nearest neighbor As (at a distance
$R=a\sqrt{3}/2=2.4\AA$), arranged in a perfect tetrahedron, two in the
upper and two in  the lower layer. (The real structure has somewhat
distorted tetrahedra).

In our model, we assume that the valence electrons occupy 3$d$
orbitals of the Fe atoms. We are interested in excess charges on the
Fe sites, be they electrons or holes, and we describe them in an
effective single band model: we assume that they can hop between
nearest neighbor Fe, with a hopping integral $t$. We will also
consider separately the effect of second nearest neighbor hopping,
with a hopping integral $t'$. There is very little hybridization of
these 3$d$ orbitals with As orbitals, and we ignore it
altogether~\cite{George}. On-site Hubbard repulsion for these charges,
characterized 
by an energy $U_H$, is also included. The value of $t, t', U_H$ and of
the other parameters introduced below is discussed in the next
subsection. Thus, the hopping Hamiltonian for the extra charges on the
Fe sites is:
\begin{equation}
\EqLabel{e1} {\cal H}_{\rm Fe}=\hat{T}+ \hat{T}' +\hat{U}= -t
\sum_{\langle i,j\rangle, \sigma}^{} \left( c_{i,\sigma}^\dagger
c_{j,\sigma} + h.c.\right)-t' \sum_{\langle\langle i,j\rangle\rangle,
\sigma}^{} \left( c_{i,\sigma}^\dagger c_{j,\sigma} + h.c.\right) +U_H
\sum_{i}^{} \hat{n}_{i\uparrow} \hat{n}_{i\downarrow}
\end{equation}
where $c^\dagger_{i,\sigma}$ creates an extra charge on the Fe site
$i$ with spin $\sigma$, $\hat{n}_{i\sigma} =
c^\dagger_{i\sigma}c_{i\sigma}$ and $\langle i, j\rangle,
\langle\langle i,j\rangle\rangle$ denote summation over the nearest,
respectively second nearest neighbor Fe sites. A more detailed model
taking in consideration the various relevant $3d$ orbitals is of
course possible, but not necessary for our purposes. We avoid such
complications in order to make more transparent the essential new
physics proposed in our model.

Ab-initio calculations find that the anions are effectively in the As$^{3-}$
state, with fully occupied  $4p$ bands lying well below the Fermi
energy~\cite{George}. In reality, 
these As $4p$ states are strongly hybridized with Fe $4s$ and $4p$ 
orbitals (this hybridization is responsible for the fact that the
effective As charge inside 
the muffin tin is much less than the ionic charge of
$-3$, ~\cite{George}). Since this mixing is strongly dependent on the Fe $3d$
occupation, it will provide a screening mechanism for the bare Coulomb
repulsion, in particular for the value of $U_H$. In the following we consider
that such screening is already included in $U_H$. For simplicity,
we also refer only to the full As $4p$ orbitals, ignoring their
hybridization with the Fe $4s$ and $4p$ orbitals (this could also be
accounted for in a multi-band model, but is of limited relevance and we
ignore it for clarity).

In our model, the  only role of these As ions is that they become
polarized due to the electric field created when an extra charge
(electron or hole) is on a
nearby Fe atom. This is due to transitions of As electrons from the
filled $4p$ shell into (primarily) the empty $5s$ shell, and results
into a dipole 
moment (polarization cloud) pointing towards the Fe atom which hosts
the extra charge. For simplicity, we only consider the polarization of
the 4 As atoms closest to the Fe hosting a charge and we ignore the
dipole-dipole interactions between these As polarization  clouds. The
quantitative importance of these approximations is discussed towards the
end of this supplementary material. 

Three more observations are in order. First, it is more convenient to
describe the As polarization clouds not in terms of an electron being
excited from the filled 4$p$ orbitals into the empty $5s$ orbital but,
equivalently, in terms of a hole being excited from the $5s$ into the
$4p$ orbitals. Thus, each As will be described by 4 
operators: $s^\dagger_\sigma, p_{x,\sigma}^\dagger,
p_{y,\sigma}^\dagger, p_{z,\sigma}^\dagger$, which 
create a hole of spin $\sigma$ in the respective orbital. The
ground-state, then, is 
$s_\uparrow^\dagger s_\downarrow^\dagger|0\rangle$ and the
$p$-orbitals lie at an energy $\Omega$ 
above it. If we measure energies above the $s$-orbital, the Hamiltonian of
the unperturbed As ions is:
\begin{equation}
\EqLabel{e2}
{\cal H}_{\rm As} = \Omega \sum_{i,\lambda,\sigma}^{}
p^\dagger_{i,\lambda,\sigma} p_{i,\lambda,\sigma}  
\end{equation}
where $i$ is the location of the As atom (see below),
$\lambda=x,y,z$ and $\sigma$ is the spin of the hole. 

The second observation is that this Hamiltonian ignores hopping of
charges between As atoms. This is a reasonable approximation given the
large distance between neighbor As atoms, which results in rather narrow As
bands, as demonstrated by ab-initio calculations. This is why we model
the As within this rather atomistic picture.

\begin{figure}[t]
\includegraphics[width=0.50\columnwidth]{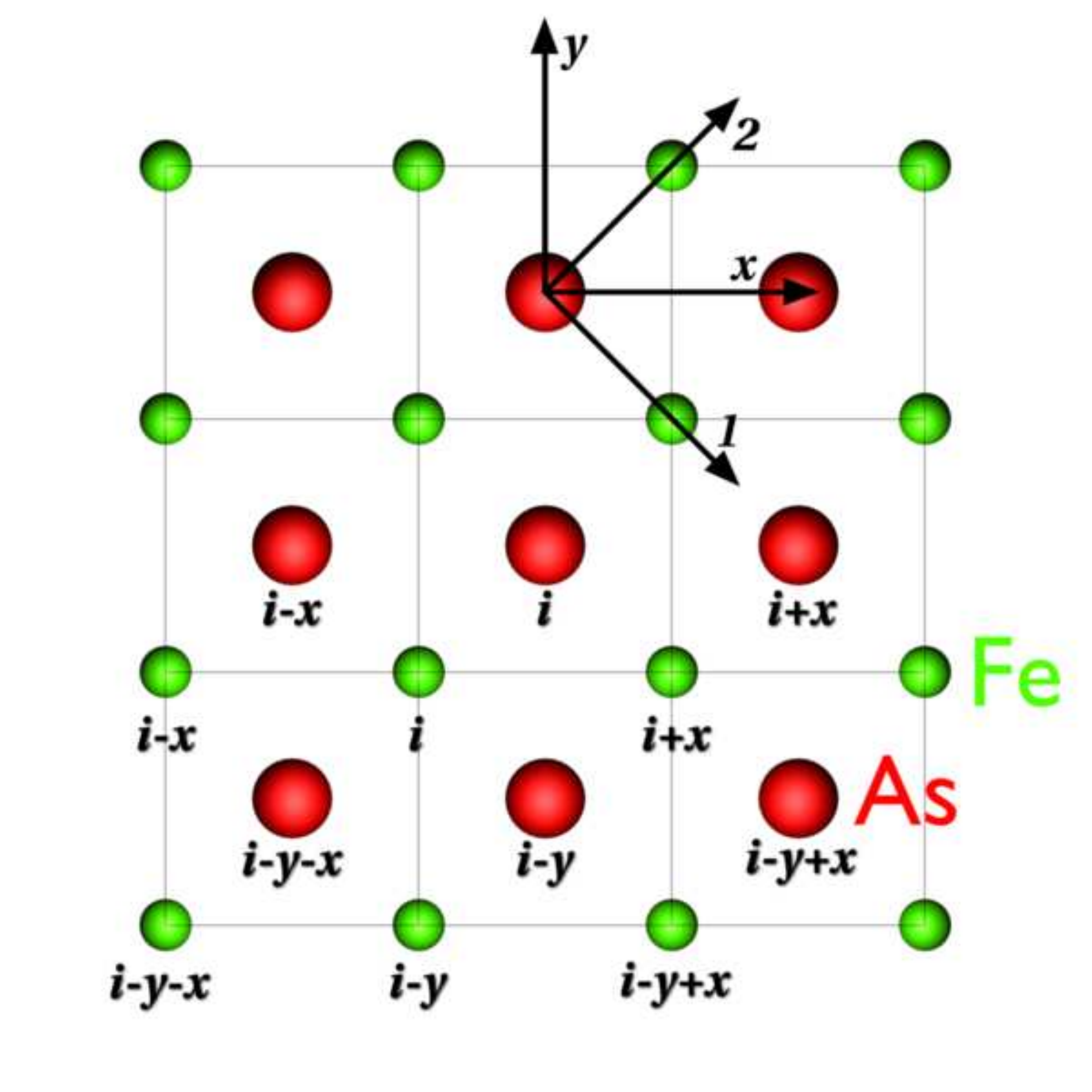}
\caption{Top view of the Fe-As structure. Several sites are
  indexed. The orientations of the in-plane axes 1,2 and $x,y$ is also
  indicated. The Fe and As layers have different depths in the $z$ direction.}
\label{fig2}
\end{figure}

Finally, the lattice illustrated in Fig. \ref{fig1} suggests that we
should consider a unit cell with (a minimum of) 2 Fe atoms and 2 As atoms,
one in the top and one in the bottom layer. However, given our
assumptions, in our model it makes no difference whether
half the As are above and half are below the Fe layer, or whether they
are all in the same plane (for example, in the layer below the plane
of Fe). Mathematically, this follows because for the As in the layer
above, one can redefine the $p_z$ operators as $p_z \rightarrow
-p_z$, which is equivalent to a local flipping of the direction of the
$z$-axis. This is very convenient, since it allows us to use a simple
unit cell with one Fe and one As in the basis, as illustrated
schematically in Fig. \ref{fig2}. However, it is
important to point out that if one wants to include, for example, As
dipole-dipole interactions, than one needs to carefully consider the
true orientations of the As polarization clouds with respect to one
another.

Figure \ref{fig2} also indicates our choice for indexing the lattice
sites, with Fe$_i$ and As$_i$ being the unit cell of the simple square
lattice. If the charge is on Fe$_i$, the polarization of As$_i$ and
the other 3 As neighbors is 
pointing in its
direction. It is therefore more  convenient to use the $1,2$ axes
orientation for 
the in-plane directions (see Fig. \ref{fig2}), and $3=z$ for the
out-of-plane direction, in terms of which $p_1 = {1\over \sqrt{2}}
( p_x - p_y), p_2 = {1\over 
  \sqrt{2}} ( p_x + p_y), p_3 = p_z$ and ${\cal H}_{\rm AS}$
remains formally identical to that of Eq. (\ref{e2}), except that now
$\lambda=1,2,3$. 

With these definitions, the cloud of As$_i$ is pointing towards Fe$_i$
 in the direction
 $-\sin\theta \vec{e}_2 + \cos \theta \vec{e}_3$, where $\cos
\theta = {1\over \sqrt{3}}$ comes from simple geometry illustrated in
 Fig. \ref{fig1}. Of course, Fe$_i$ also interacts
with  As$_{i-x}$, As$_{i-x-y}$ and As$_{i-y}$. Their polarization vectors
are obtained by appropriate rotations from the one just listed. 

The interaction Hamiltonian is, then:
$$ {\cal H}_{\rm int} = g \sum_{i, \sigma}^{} \hat{n}_i
\left[s_{i,\sigma}^\dagger\left(-\sin \theta p_{i,2,\sigma} + \cos
\theta p_{i,3,\sigma}\right)+ s_{i-x,\sigma}^\dagger\left(\sin \theta
p_{i-x,1,\sigma} + \cos \theta p_{i-x,3,\sigma}\right)\right.
$$
\begin{equation}
\EqLabel{1} \left.+ s_{i-x-y,\sigma}^\dagger\left(\sin \theta
p_{i-x-y,2,\sigma} + \cos \theta p_{i-x-y,3,\sigma}\right)+
s_{i-y,\sigma}^\dagger\left(-\sin \theta p_{i-y,1,\sigma} + \cos
\theta p_{i-y,3,\sigma}\right) + h.c. \right]
\end{equation}
where the sum is over all unit cells $i$ in the lattice. This
interaction 
describes the $s\rightarrow p$ transitions and resulting polarization
of the As clouds (with the appropriate orientations) if extra charges,
counted  by $\hat{n}_i = \hat{n}_{i\uparrow} + \hat{n}_{i\downarrow}$,
are on Fe$_i$ site. 

Our model Hamiltonian ${\cal H} = {\cal H}_{\rm Fe} + {\cal H}_{\rm As} +
{\cal H}_{\rm int}$ is thus characterized by 5 parameters, $t, t',
U_H, \Omega,
g$. We now discuss the choice of their values.

\subsection{Parameters}

Ab-initio calculations find the  Fe 3$d$ dispersional part of the
bandwidth to be  roughly $W=2eV$~\cite{George}. 
Our 2D hopping model has a  bandwidth $W = 8t$ irrespective of the
value of $t'$,
thus  $t=0.25$eV is a typical value. For $t'$, we will consider both
the case $t'=0$ and $t'=-t/2=-0.125eV$, which is the appropriate
value for hopping between $d_{xy}$ and the like orbitals  which
contribute most 
near the Fermi level. The Hubbard repulsion $U_H$ is certainly a large
energy. We will use it as a parameter, and show that our results are
very weakly sensitive to its precise value once $U_H > 8$ eV or so,
which is certain to be the case. 

The ab-initio calculations also suggest a difference from the top of
the As $4p$ bands to the bottom of the As $5s$ bands of about
4eV. This value provides a lower limit for the split $\Omega$ between
the $s$ and $p$ orbitals. A more appropriate measure
would be the distance between the mid-points of these bands, but this
is harder to estimate. We use  $\Omega = 6$eV as a
typical value, and  also study the effects of  larger
$\Omega$ values on the results. This is also reasonable
because higher-energy orbitals of the As will be polarized as well, so
one can think of these $s$ and $p$ orbitals as modeling effectively
the polarization of the whole atom.

The final parameter needed is $g$. Since this characterizes the $s-p$
hybridization when an As polarization cloud is created, we should extract it
from the As polarizability $\alpha_p$. Its value is not
known precisely, but the polarizability is typically equal to the
volume of the atom. Since  As is a big atom, its polarizability
is estimated to be $\alpha_p = 10-12 \AA^3$.

Consider the polarization cloud formed on an As ion, due to a charge $q$
placed at a distance $R$ away, in the direction characterized by the
versor $\vec{e}$. The electric field at the As site is
then $\vec{E} = 
{q\over R^2} \vec{e}$. Note that we use an unscreened value for this
electric field. The reason is that we are interested in behavior
related to the
high-frequency 
dynamic electronic polarizability, and not that due to slow
vibrational motion of the As themselves. As we show
below, the effective masses of the polarons and bipolarons are fairly
small, meaning that they are highly mobile. The As ions are fast to
acquire a polarization cloud, but since they are very heavy,  they simply
cannot move fast enough to follow the fast polaron dynamic.  This
separation of energy scales allows us to ignore lattice dynamics
effects in
the following. The interaction between the As and the electric field
is  ${\cal H}_{\rm int}= -
\hat{\vec{p}}\cdot \vec{E}$, where $\hat{\vec{p}}$ is the As dipole
moment operator. In the second quantization, and if we restrict
ourselves only to $s$ and $p$ orbitals to describe the As, its
dipole moment is
\begin{equation}
\EqLabel{e3}
\hat{\vec{p}}=\sum_{\lambda,\sigma} \langle s | e \vec{r}|
p_\lambda\rangle \left( s_\sigma^\dagger p_{\lambda,\sigma} + h.c.\right)
\end{equation}
where $\lambda=x,y,z$ or $\lambda=1,2,3$ (either basis is
equivalent), and we already took in consideration the fact that there are
no matrix elements of the dipole moment between orbitals of the same
level ($\langle s |\vec{r}| s\rangle=0$, etc). Moreover, $\langle s |
e \vec{r}| 
p_\lambda\rangle =  e \vec{e}_\lambda a_{\rm As}$, since the operator $\vec{r}$
has a non-vanishing matrix element only in the direction
$\vec{e}_\lambda$ of the
orbital $p_\lambda$ involved. Thus, $a_{\rm As}$ is a characteristic
``size'' of 
the ion, defined as:
\begin{equation}
\EqLabel{e4}
a_{\rm As} = \langle s | x | p_x\rangle = \int_{}^{} d \vec{r}
\phi_x^*(\vec{r}) x \phi_{p_x}(\vec{r})= \langle s | y | p_y\rangle
=\langle s | z | p_z\rangle  
\end{equation}
In terms of this, we find:
\begin{equation}
\EqLabel{e5}
{\cal H}_{\rm int}=  -{ea_{\rm As}q\over R^2} \sum_{\lambda, \sigma}^{}
\left(\vec{e}\cdot \vec{e}_\lambda s_{\sigma}^\dagger p_{\lambda,\sigma}
+ h.c.\right)
\end{equation}
This is exactly the form used in Eq. (\ref{1}), except  there we
have explicitly written the various projections $\vec{e}\cdot
\vec{e}_\lambda $ in terms of the angle $ \theta$ appropriate for our
geometry. It follows that $g =  -{ea_{\rm As}q/ R^2}$. As we show in the
following, the energetics only depends on $g^2$, i.e. it is the same
for both electron or hole extra charges on the Fe atom. Since the sign
of $g$ is irrelevant, from now on we assume that the extra charges are
electrons, $q=-e$, so that: 
\begin{equation}
\EqLabel{e6}
g =  {a_{\rm As}e^2\over R^2}>0.
\end{equation}
This shows that $g$ is an energy equal to the typical induced  dipole
$ea_{\rm As}$ times the applied electric field $e/R^2$.

For simplicity, in the rest of this discussion let us denote $p_\sigma =
\sum_{\lambda}^{} \vec{e}\cdot\vec{e}_\lambda 
p_{\lambda,\sigma}$, 
i.e. it is the annihilation operator for the $p$ orbital oriented in
the direction $\vec{e}$ of the applied electric field. Of course, one
can form two other linear 
combinations of the $p_x, p_y, p_z$ orbitals which are orthogonal to
this direction. These two do not couple to the electric field
and do not participate in the polarization cloud, thus their
energetics is trivial. The interesting part for the problem of the As
cloud in the presence of the extra charge is therefore described by
the Hamiltonian: 
\begin{equation}
\EqLabel{e7}
\hat{h} = \Omega \sum_{\sigma}^{} p^\dagger_\sigma p_\sigma + g
\sum_{\sigma}^{ }
(s_\sigma^\dagger p_{\sigma} + h.c.) 
\end{equation}
This Hamiltonian is trivial to diagonalize. Its ground state has
the energy
\begin{equation}
\EqLabel{e7b}
E_{\rm cloud} = \Omega - \sqrt{\Omega^2+4g^2}
\end{equation}
and the eigenstate
\begin{equation}
\EqLabel{e8}
|GS\rangle =  \gamma^\dagger_\uparrow \gamma^\dagger_\downarrow |0\rangle
\end{equation}
where
\begin{equation}
\EqLabel{e9}
\gamma_{\sigma}^\dagger = \cos \alpha s_\sigma^\dagger - \sin \alpha
p_\sigma^\dagger
\end{equation}
and 
\begin{eqnarray}
\nonumber \cos \alpha &&= \sqrt{{1\over 2}\left(1+{\Omega\over
	\sqrt{\Omega^2+4g^2}}\right)}\\
\EqLabel{e10} \sin \alpha  &&= \sqrt{{1\over 2}\left(1-{\Omega\over
	\sqrt{\Omega^2+4g^2}}\right)}
\end{eqnarray}
Indeed, note that if $g=0$ this is reduced to the As ground state
(holes in the $s$ orbitals, $\alpha=0^o$) whereas for $g \rightarrow
\infty$, the 
$\gamma$ operators describe the maximally polarized linear combination 
$(s-p_x)/\sqrt{2}$ and $\alpha=45^o$.

The second eigenstate of $\hat{h}$ has an energy $E_{\rm exc} = \Omega +
\sqrt{\Omega^2+4g^2}$ and corresponds to filling up the orbitals
$\eta_\sigma^\dagger = \sin \alpha s_\sigma^\dagger +\cos \alpha
p_\sigma^\dagger$, i.e. of creating a polarization cloud oriented
antiparallel to the electric field. As already discussed, one can also
excite the electron in the $p$ orbitals orthogonal to the direction of
the electric field, the energy of those states being simply
$\Omega$. Given that $\Omega$ is a large energy, all these excited states
are well above the ground state and will be ignored in the rest of the
calculation.

Finally, the  dipole moment induced on the As atom is:
\begin{equation}
\EqLabel{e11}
\langle p \rangle = \langle GS| \vec{e}\cdot \hat{\vec{p}}| GS \rangle =
4ea_{\rm As} \sin \alpha \cos \alpha = {4ea_{\rm As}g\over
  \sqrt{\Omega^2+ 4g^2}} 
\end{equation} 
Remember that Eq. (\ref{e6}) revealed that $g$ is proportional to the
applied electric field $E=e/R^2$. This shows  that in the limit $g\ll
\Omega$ we can approximate 
$$
\langle p \rangle \approx {4ea_{\rm As}g\over \Omega} = { 4 e^2 a_{\rm
	As}^2\over \Omega} E 
$$
and thus identify the polarizability from this linear regime where
$\langle p \rangle = \alpha_p E$. This allows us to extract $a_{\rm
  As}$ in terms 
of $\alpha_p$ and then, through Eq. (\ref{e6}), to calculate:
\begin{equation}
\EqLabel{e12}
g^2 = {\alpha_p \Omega e^2\over 4 R^4} 
\end{equation}
Using $\alpha_p = 10\AA^3$ and $\Omega = 6$eV gives the estimate $g =
2.5$eV and thus $g/\Omega =0.4$, suggesting that non-linear effects
described by our model are starting to become important, however they
are not dominant yet. For example, with these numbers, the energy of the
polarized As cloud  from Eq. (\ref{e7b}) is $E_{\rm cloud} = -1.78$ eV,
whereas in the 
linear regime $g/\Omega \ll 1$ we would have
$E_{\rm cloud} \approx -{2g^2\over \Omega} =-2.05$eV.
As a consistency check, also note that the linear expression for the
energy of the cloud can be rewritten as $E_{\rm cloud} \approx
-{2g^2\over \Omega} =-{\alpha_p E^2\over
2}$, as expected to be the case in this regime.

\begin{figure}[t]
\includegraphics[width=0.50\columnwidth]{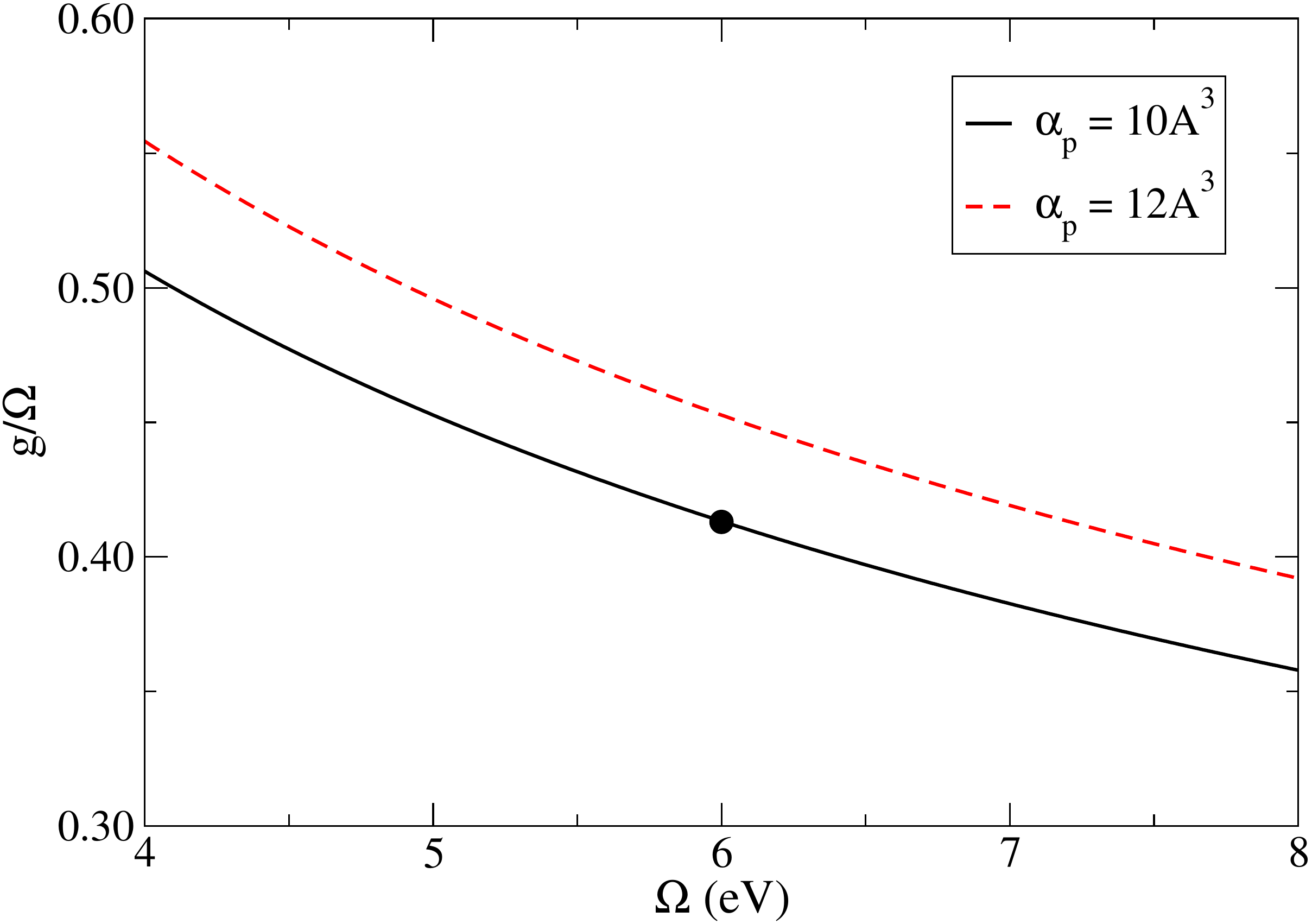}
\caption{Ratio $g/\Omega$ vs. $\Omega$, for a fixed polarizability
  $\alpha_p=10\AA^3$ (full line) and 12$\AA^3$ (dashed line). The dot
  indicates our typical choice of parameters.}
\label{fig3}
\end{figure}

To conclude,   we will use $t=0.25$eV, $t'=0$ or
-0.125eV, $\Omega = 6$eV
and $g=2.5$eV 
as typical numbers. As discussed previously, we will also allow
$\Omega$ to vary and we will then choose $g$ according to Eq. (\ref{e12}),
so that the polarizability stays constant. The corresponding
$g/\Omega$ ratio is shown in Fig. \ref{fig3}: for a wide range of
values of $\Omega$, $g/\Omega$ is roughly 0.4-0.5.

\section{The single electronic polaron}

We would like to find the eigenstates of the total Hamiltonian ${\cal
H} = {\cal H}_{\rm Fe} + {\cal H}_{\rm As} + {\cal H}_{\rm int}$ in the case
when there is a single extra charge  at one of the Fe
sites. Of course, this charge can hop around and will also polarize
 the As atoms in its vicinity. We call this composite object (the
 charge dressed by the 
surrounding polarization clouds) an electronic polaron.

To find its low-energy spectrum, we will use
perturbation theory in the hopping Hamiltonian. The reason this is a valid
approach is that $t$ is by far the smallest energy scale in this
problem. A more detailed
justification is provided below. 

\subsection{Zero order perturbation: $t=0$}

In the absence of hopping the electron will always stay on the same
site, say at Fe$_i$, and  will polarize its 4 neighbor As. Since 
we ignore dipole-dipole interactions in our model, this simply leads
to the appearance of 4 polarization clouds like
the one discussed in the previous section,  each  pointing
towards the Fe 
atom. This is illustrated in Fig. \ref{fig3}. 

\begin{figure}[t]
\includegraphics[width=0.30\columnwidth]{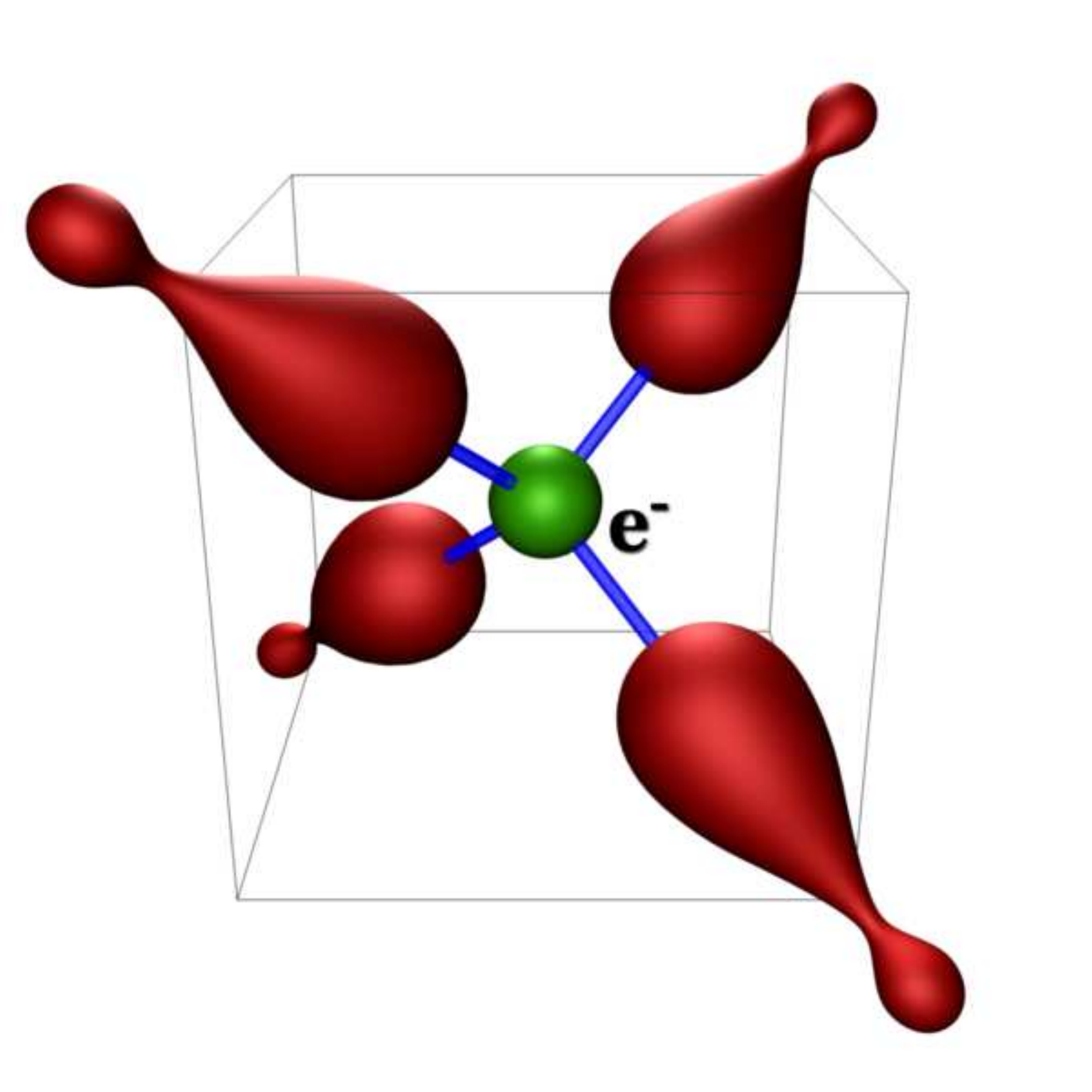}
\caption{Pictorial illustration of a static electronic polaron. The
  electron residing on the central Fe atom polarizes the holes on the
  nn As sites into the $4p$ orbital pointing towards the Fe. More
  distant As are assumed to be in the $5s$ ground-state. }
\label{fig4}
\end{figure}

The ground-state energy of the static polaron is then, immediately:
\begin{equation}
\EqLabel{p1}
E_{P,GS} = 4 E_{\rm cloud} = 4 \left( \Omega - \sqrt{\Omega^2+ 4g^2}\right).
\end{equation}
The lowest excited state is at an energy ${1\over 2} \left[ \Omega +
  \sqrt{\Omega^2+ 4g^2}\right]$ 
above $E_{P,GS}$, and  has either the spin-up or spin-down hole of one
of the As 
atoms polarized in a 
direction perpendicular to its local electric field. These excited
states will be ignored in the following, since $t\ll \Omega$. 

To characterize the polarization clouds, we  introduce the following
operators  [analogs of Eq. (\ref{e9}) for the specific
geometry here]:
\begin{equation}
\EqLabel{p2}
\gamma_{i,\lambda,\pm,\sigma}^\dagger = \cos \alpha
s_{i,\sigma}^\dagger - \sin \alpha
\left(\pm \sin \theta p_{i,\lambda,\sigma}^\dagger + \cos \theta
p_{i,3,\sigma}^\dagger\right) 
\end{equation}
where here $\lambda=1,2$ shows if the in-plane component of the
polarization cloud is oriented along axis 1 or axis 2, and the $\pm$
sign indicates whether the in-plane polarization is parallel or
antiparallel to this axis. The angles $\alpha$ and $\theta$ were
defined in the previous sections.

In terms of these, the ground-state of the system with the electron at
site Fe$_i$ is:
\begin{equation}
\EqLabel{p3}
|\Phi_i\rangle = c^\dagger_i|i\rangle
\end{equation}
where $|i\rangle$ describes the state of all As atoms when the electron
is at site Fe$_i$ (since there is a single such electron, its
spin is irrelevant and we drop its index here). Following our
discussion in the previous section, we have
\begin{equation}
\EqLabel{p3b}
|i\rangle = 
\prod_{\sigma}^{}
\gamma^\dagger_{i,2,-,\sigma}\gamma^\dagger_{i-x,1,+,\sigma}
\gamma^\dagger_{i-x-y,2,+,\sigma}\gamma^\dagger_{i-y,1,-,\sigma}|GS\rangle_i
\end{equation}
where 
\begin{equation}
\EqLabel{p4}
|GS\rangle_i = \prod_{|j-i|>R,\sigma}^{} s^\dagger_{j,\sigma}|0\rangle,
\end{equation}
i.e. it describes unpolarized (ground-state) As atoms at all 
sites which are 
not nearest neighbor to the site Fe$_i$ containing the extra
charge. Thus, the product of
$\gamma$ operators describes the polarization clouds on the 4 neighbor
As sites (both spin-up and spin-down holes are polarized), and
$|GS\rangle_i$ describes the remaining holes on the unpolarized As sites.
Of course, this ground-state $|\Phi_i\rangle $ is highly degenerate
since the electron 
can be at any Fe site. The hopping $t$ lifts this degeneracy.

\subsection{First order perturbation}

In the subspace generated by all the ground-states $|\Phi_i\rangle $
described in 
the previous section, we can form eigenstates which are invariant to
translations in the 
usual fashion:
\begin{equation}
\EqLabel{p5}
|\Phi_{\vec{k}}\rangle = \sum_{i}^{} {e^{i\vec{k}\cdot
  \vec{R}_i} \over N} |\Phi_i\rangle 
\end{equation}
where the lattice has $N\times N=N^2$ unit cells  and we assume
periodic boundary 
conditions. As usual, the quasi-momenta $\vec{k}$ are defined inside the
first Brillouin zone of the simple square lattice $-{\pi \over
  a} < k_x, k_y \le {\pi\over a}$.

\begin{figure}[t]
\includegraphics[width=0.50\columnwidth]{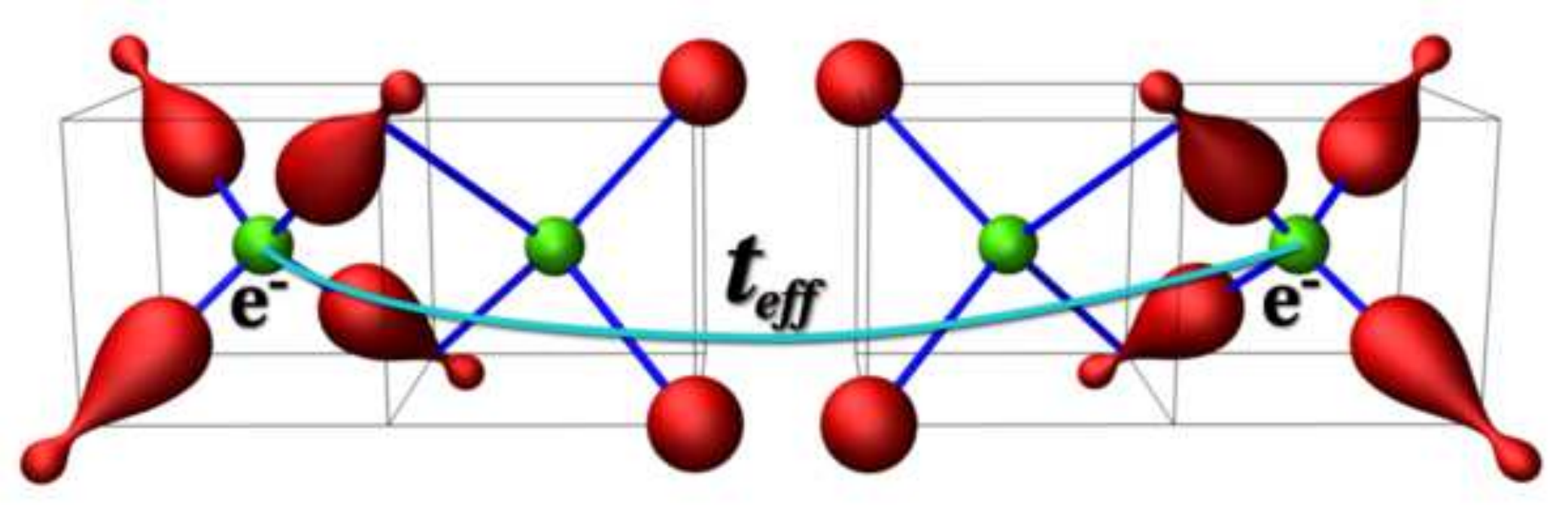}
\caption{Single polaron hopping in the $x$-direction. The hopping integral
is renormalized by the overlap between the polarization clouds in the
initial and the final position. The unpolarized As atoms are in their
ground-state $s$-orbital.}
\label{fig5}
\end{figure}

Since the hopping Hamiltonian is diagonal in this basis,
$\langle \Phi_{\vec{k}}| \hat{T}_{\rm tot} |  \Phi_{\vec{k}'}\rangle
\propto \delta_{\vec{k},\vec{k}'}$, it follows that
$|\Phi_{\vec{k}}\rangle$ are the
eigenstates in first-order perturbation theory. Their 
energies are:
\begin{equation}
\EqLabel{p6}
E_{P}(\vec{k}) = E_{P,GS} + \langle \Phi_{\vec{k}}| \hat{T}_{\rm tot} |
\Phi_{\vec{k}}\rangle 
\end{equation}
with the second term describing the polaron
dispersion. Straightforward calculation reveals that: 
\begin{equation}
\EqLabel{p7}
E_{P}(\vec{k}) = E_{P,GS} - 2 t_{\rm eff} \left[ \cos (k_x a) + \cos
(k_ya)\right] -4t'_{\rm eff}\cos (k_x a) \cos (k_y a)=E_{P,GS}
+\epsilon_{\rm eff}(\vec{k}), 
\end{equation}
in other words the dispersion is identical to that of a free particle,
but with 
renormalized hopping integrals (or mass). The renormalization is due to
the overlap of the corresponding As clouds as the electron moves from
one site to a nearest or second-nearest neighbor site, as illustrated
schematically in 
Fig. \ref{fig5} for the former case. These overlaps result in:
\begin{equation}
\EqLabel{p8}
t_{\rm eff} = t \langle i| i+x\rangle = t \left[{1\over 6} \left(1+{\Omega\over
  \sqrt{\Omega^2 + 4g^2}}\right)\left(2+{\Omega\over
  \sqrt{\Omega^2 + 4g^2}}\right)\right]^4
\end{equation}
\begin{equation}
\EqLabel{p8b}
t'_{\rm eff} = t' \langle i| i+x+y\rangle = t'\left[{1\over 2}
  \left(1+{\Omega\over  \sqrt{\Omega^2 + 4g^2}}\right)\right]^{6}
\left[{1\over 3} \left(1+{2\Omega\over 
  \sqrt{\Omega^2 + 4g^2}}\right) \right]^2 
\end{equation}

\begin{figure}[b]
\includegraphics[width=0.50\columnwidth]{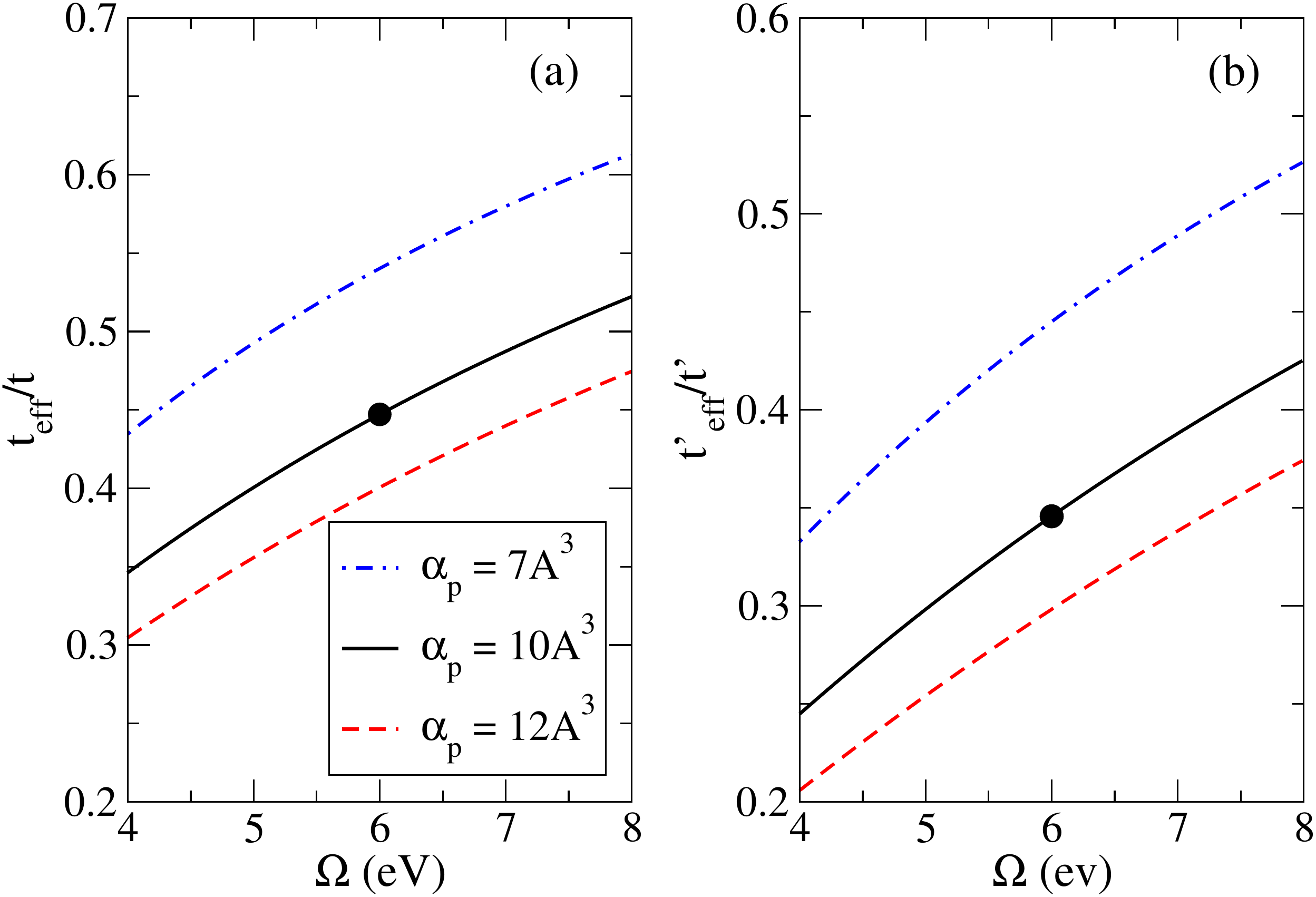}
\caption{$t_{\rm eff}/t$ (left) and $t'_{\rm eff}/t'$ (right)
  vs. $\Omega$, for different As 
  polarizabilities. The dots indicate the main values used here.}
\label{fig6}
\end{figure}

The results are plotted (in units of bare $t$, respectively $t'$) in
Fig. \ref{fig6}, 
vs. $\Omega$, for the two values of polarizability $\alpha_p=10$ and
$12\AA^3$. As expected, in the limit of large $\Omega$, which implies
$g/\Omega \rightarrow 0$, the ratios increase towards unity because
the polarization clouds become very small. As $\Omega$ decreases, the
effective coupling $g/\Omega$ increases and the ratios (and therefore
the bandwidths) decrease. This is
typical polaronic behavior, however note
that even if we are in the strong coupling limit $g/\Omega \rightarrow
\infty$, the effective hopping is lowered to at most $t_{\rm eff}/t =
1/3^4=0.012, t'_{\rm eff}/t'= 1/(2^6 3^2)=0.0017$. While these are, of
course, small values, the renormalized hoppings are not becoming 
exponentially small with increased coupling, as is the case for
polarons with phonon clouds. The reason is that within this model, the
polarization clouds saturate to a maximum possible value and therefore the
overlap cannot become arbitrarily small. 

The effective polaron mass renormalization is inversely
proportional to the $t_{\rm eff}/t$ ratio (this statement is true if
$t'=0$. If $t'$ is finite both ratios determine the new effective
mass, however they are roughly equal). Thus, these polarons remain
rather light objects, with a mass up to a factor of 3 or less larger than
the bare band mass, if $\Omega$ is not too small.  The dots in
Fig. \ref{fig6} indicate our chosen 
values, for which the polaron mass is roughly 2.2 times the mass of
the free band electron, and $t_{\rm eff} = 0.45\times t\approx
0.11$eV.

While not huge, this renormalization of the effective
hoppings/mass is important since it further supports our use of the
perturbation theory. First order perturbation is reasonable only
assuming that half the 
effective bandwidth is much smaller than the distance to the next
set of excited states, which were ignored in this calculation. As already
discussed, those excited states are at ${1\over 2} \left[
  \Omega+\sqrt{\Omega^2+ 
  4g^2}\right]\approx 5 $eV and
higher energies,
whereas $4t_{\rm eff}\approx 0.45$eV. These numbers suggest that first
order perturbation theory should be quite accurate for parameters in
the regime of interest to us (the higher the $\Omega$ value used, the
more accurate the perturbation theory is). 
A second order calculation (which is possible, but rather involved)
should only bring in relatively small corrections. Moreover, those
corrections may well be of the same order as other ignored small terms (like
dipole-dipole interactions between the As clouds) and therefore one
would need to reconsider all these approximations carefully before
going to a higher order in perturbation theory.

In conclusion, our calculation shows that, to first order, the only
effect of the formation of the electronic polarons is to scale down the
hopping energy scales by a factor of around 2.5 (with slight differences
for nearest vs. 2nd nearest neighbor hopping), but otherwise the
dispersion is similar to that  predicted
by LDA for a free charge. ARPES measurements indicating precisely such
behavior (in LaOFeP) have 
been reported already in Ref. \onlinecite{ARPES}. In our theory, ARPES
would measure a quasiparticle peak at low energies, with a
quasiparticle weight
\begin{equation}
\EqLabel{p15}
Z= |\langle \Phi_{\vec{k}}|c^\dagger_{\vec{k}}|GS\rangle|^2 =
\left[\cos \alpha\right]^{16}
\end{equation}
which is independent of momentum (second and higher order corrections
to the wavefunctions may bring in some small momentum dependence). For
our typical values, this implies $Z=0.38$. The remaining spectral
weight should be observed at much high energies typical of the
excited cloud states, of the order ${1\over 2} \left[ \Omega+
  \sqrt{\Omega^2 + 4g^2}\right] =
5eV$ or more.

\section{The bipolaron}

Within the same model and with the same approximations (first-order
perturbation 
theory) we will now calculate the eigenstates of the system when 2
charges are present on the Fe sublattice. In particular, we are
interested to see whether they become bound into a so-called
bipolaron, or a state of two free polarons is energetically more
favorable.

\subsection{Zero order perturbation: interaction energies}

All the eigenstates of the two charges must be either a singlet or a
triplet in the spin space. Since we expect a singlet to be the
ground-state (this 
expectation is verified by our calculations) we focus on calculations
for singlet eigenstates. Most of the calculations for triplets are
very similar, therefore we will only point out where there are major
differences between the two cases. 

For convenience, we introduce a singlet creation operator, defined as:
\begin{equation}
\EqLabel{b1}
s^\dagger_{i,j} = 
\left\{
\begin{array}[c]{cc}
{1\over \sqrt{2}} \left(c^\dagger_{i\uparrow}
c^\dagger_{j\downarrow} - c^\dagger_{i\downarrow}
c^\dagger_{j\uparrow} \right) & \mbox{ , if } i\ne j,\\
c^\dagger_{i\uparrow} 
c^\dagger_{i\downarrow} & \mbox{ , if } i= j.\\
\end{array}
\right.
\end{equation}

Within zero-order perturbation theory ($t=0$) the charges cannot move,
being pinned at their initial locations. Depending
on the distance between them, different types of polarization
clouds can form with different energies. We will now calculate these energies.

If the two electrons are 3rd nearest neighbors or further apart,
$|i-j|\ge 2a$, then their polarization clouds do not overlap (there is
no As atom that is a nearest neighbor for both of the charged
Fe sites). Therefore each charge forms the 
same 4 clouds as described for single polarons, and the energy is
twice the energy of a polaron:
\begin{equation}
\EqLabel{b2}
E_{BP,\infty} = 2\times E_{P,GS}= 8\left[\Omega - \sqrt{\Omega^2+4g^2}\right]
\end{equation}

\begin{figure}[t]
\includegraphics[width=0.50\columnwidth]{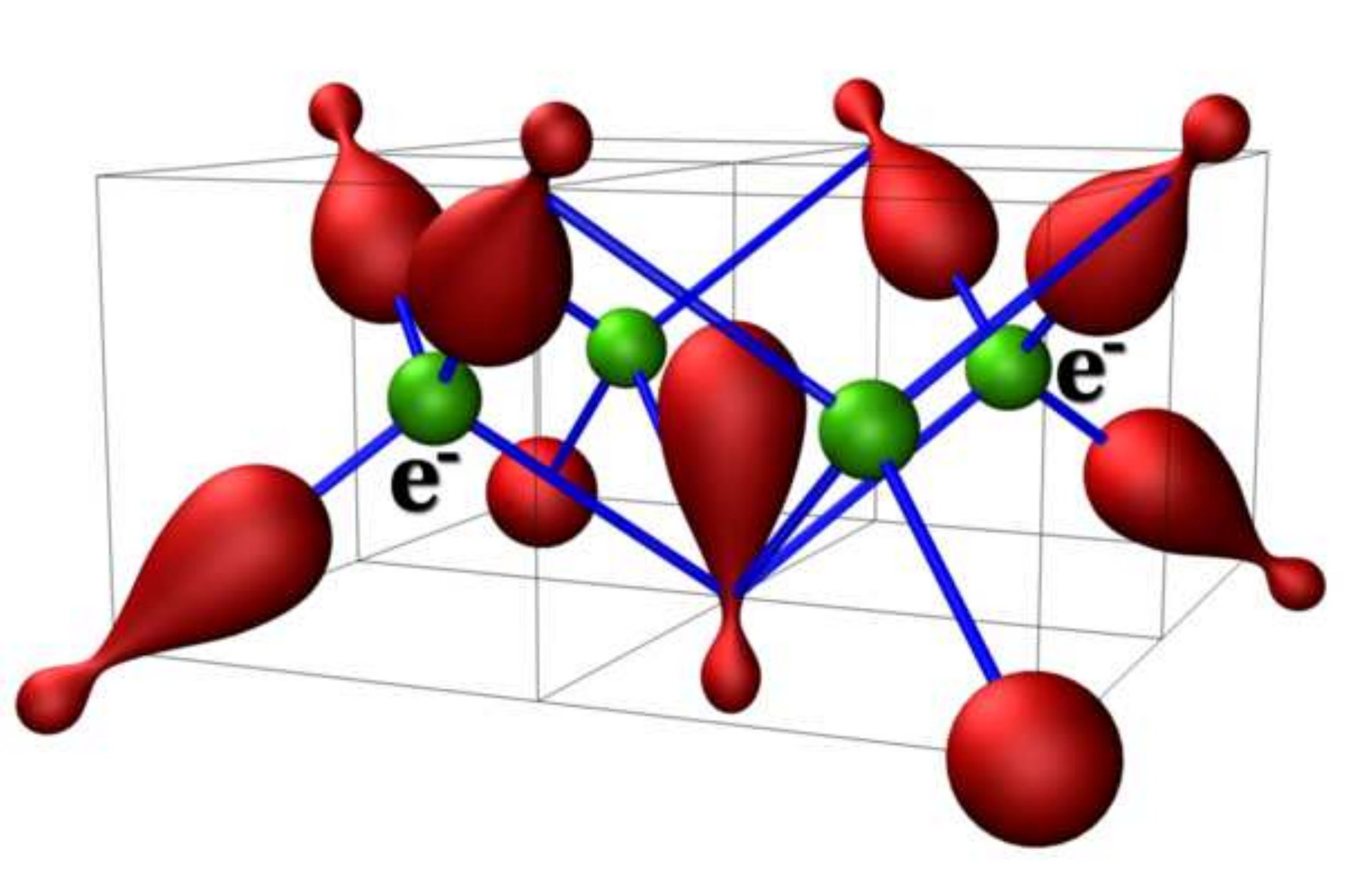}
\caption{Polarization clouds for a 2nd nearest neighbor electronic
  bipolaron. The  central ``shared'' As atom has a polarization
  different from that of the usual polaron clouds. }
\label{fig7}
\end{figure}

If the electrons are 2nd nearest neighbors, they share one As atom, as
shown in Fig. \ref{fig7}, so only 7 polarization clouds are
formed. Six of these are precisely like the ones discussed previously,
since each of those As atoms is in the field created by a single
charge. The cloud on the central As atom, however, is different, since
it is created by the sum of the 
electric fields from the two charges. Given the geometry of the
system, this new cloud is polarized
along the $z$-axis, and this $z$-axis polarization is twice as big because
two charges contribute to it. Thus, the only change in calculating the
energy and eigenstate of this cloud is to replace $g \rightarrow
2g\cos \theta=2g/\sqrt{3}$ in 
Eq. (\ref{e7}), and of course the generic $p$ orbital is now the
$p_z=p_3$ orbital. As  result, the energy of this cloud is changed to
$\Omega - \sqrt{\Omega^2 + {16\over 3} g^2}$ and therefore the total
energy of the 2nd nearest neighbor static bipolaron is:
\begin{equation}
\EqLabel{b3}
E_{BP,2} = 6\left[\Omega - \sqrt{\Omega^2+4g^2}\right] + \left[\Omega
  - \sqrt{\Omega^2 + {16\over 3} g^2}\right] 
\end{equation}
This energy is different from the energy of two free polarons by an
amount
\begin{equation}
\EqLabel{b4}
U_2 = E_{BP,2} -E_{BP,\infty} = 2 \sqrt{\Omega^2+4g^2}- \Omega -
\sqrt{\Omega^2 + {16\over 3} g^2} 
\end{equation}

\begin{figure}[t]
\includegraphics[width=0.5\columnwidth]{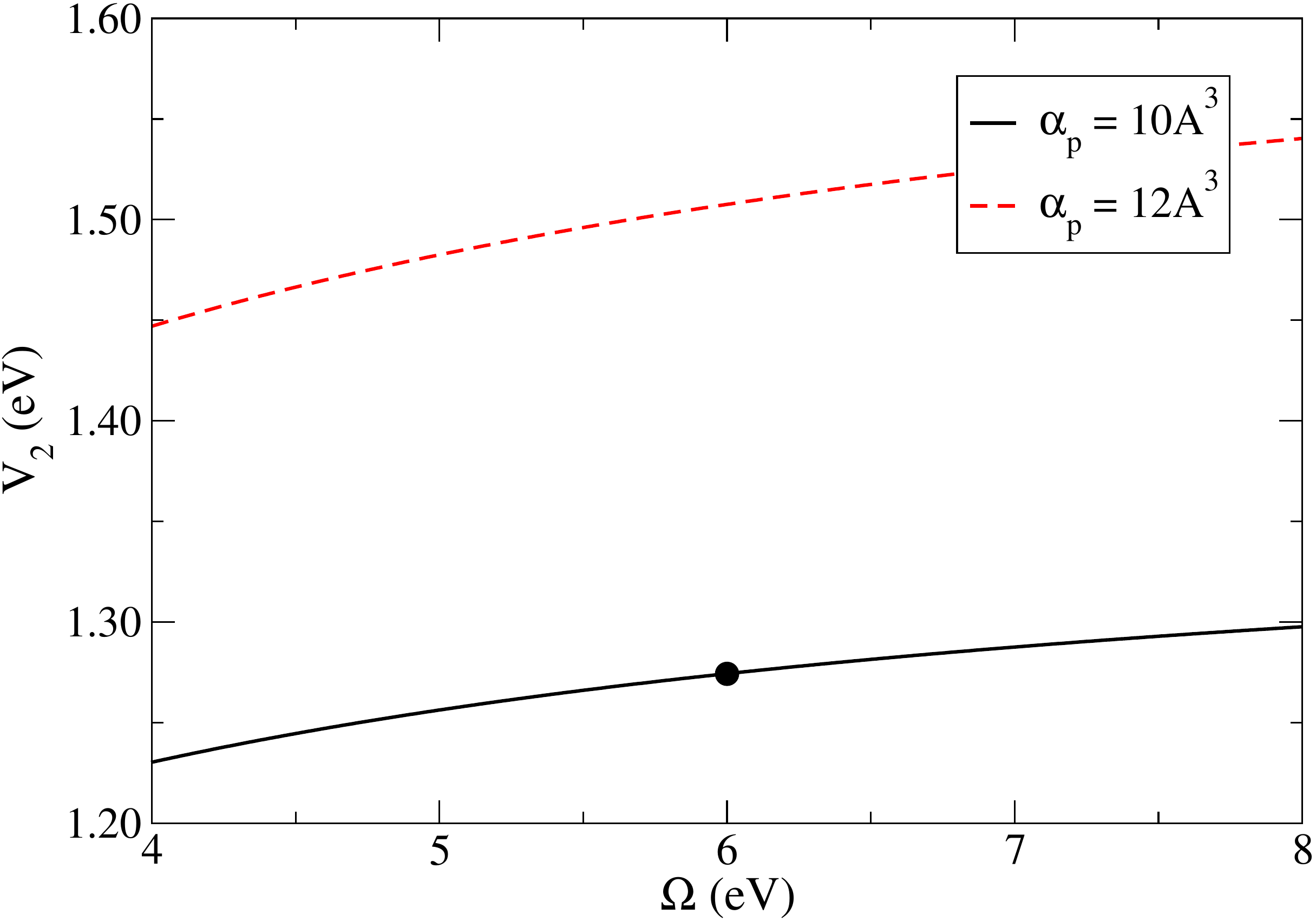}
\caption{$U_2$ vs $\Omega$, for two values of the
  polarizability. Because $U_2>0$ everywhere, it 
  follows that it is energetically more favorable for electrons to
  stay at distances $|i-j| >2$ than to be second
  nearest-neighbors. The dot indicates our typical value.}
\label{fig8}
\end{figure}

%

This is plotted in Fig. \ref{fig8} as a function of $\Omega$, for the
two values of As polarizability. In both cases, it is positive and a
slowly varying 
function of $\Omega$. The 2nd nearest neighbor effective
interaction is repulsive because the new cloud,
although bigger than a single original cloud, is not bigger than two
such clouds. As a result, it is energetically more favorable to have
two free electronic polarons situated far apart, than to have a second
nearest-neighbor bipolaron. This can be shown to be true for any value of the
polarizability and of $\Omega$ ({\em i.e} any $g/\Omega$ ratio), for
this lattice structure. 

Interestingly, this interaction can be made attractive by changing the
lattice structure. Focusing on the weak-coupling regime of small
$g\over \Omega$ and using a linear approximation, the energy of the
new cloud is $\Omega - \sqrt{\Omega^2 + 16 g^2 \cos^2\theta}\approx
-8\cos^2\theta g^2/\Omega$, while the energy of two of the original
clouds is $\approx -4 g^2/\Omega$. The 2nd nearest neighbor pair
becomes energetically favorable when $2\cos^2\theta > 1\rightarrow
\theta < 45^o$. This can be achieved by making the lattice
tetragonal, {\em i.e.} by increasing the distance between the Fe and
As layers by a factor of  $\sqrt{2}$ or more. Of course, this would
also have the effect of decreasing all the polarization energies,
since the electric 
fields would be smaller due to the increased Fe-As distances

The ground-state eigenstate of this new cloud can be obtained
from the general formula, after performing the substitutions discussed
above. This leads to the appearance of a new mixing angle:
\begin{eqnarray}
\nonumber \cos \beta &&= \sqrt{{1\over 2}\left(1+{\Omega\over
	\sqrt{\Omega^2+{16\over 3}g^2}}\right)}\\
\EqLabel{b5} \sin \beta &&= \sqrt{{1\over 2}\left(1-{\Omega\over
	\sqrt{\Omega^2+{16\over 3}g^2}}\right)}
\end{eqnarray}
which appears in the definition of the new cloud. For instance,
assuming that this new cloud appears at As$_i$ site, then its creation
operator for a given spin is:
\begin{equation}
\EqLabel{b6}
\tilde{\gamma}^\dagger_{i,\sigma} = \cos \beta s_{i,\sigma}^\dagger -
\sin \beta p^\dagger_{i,3,\sigma} 
\end{equation}

\begin{figure}[t]
\includegraphics[width=0.50\columnwidth]{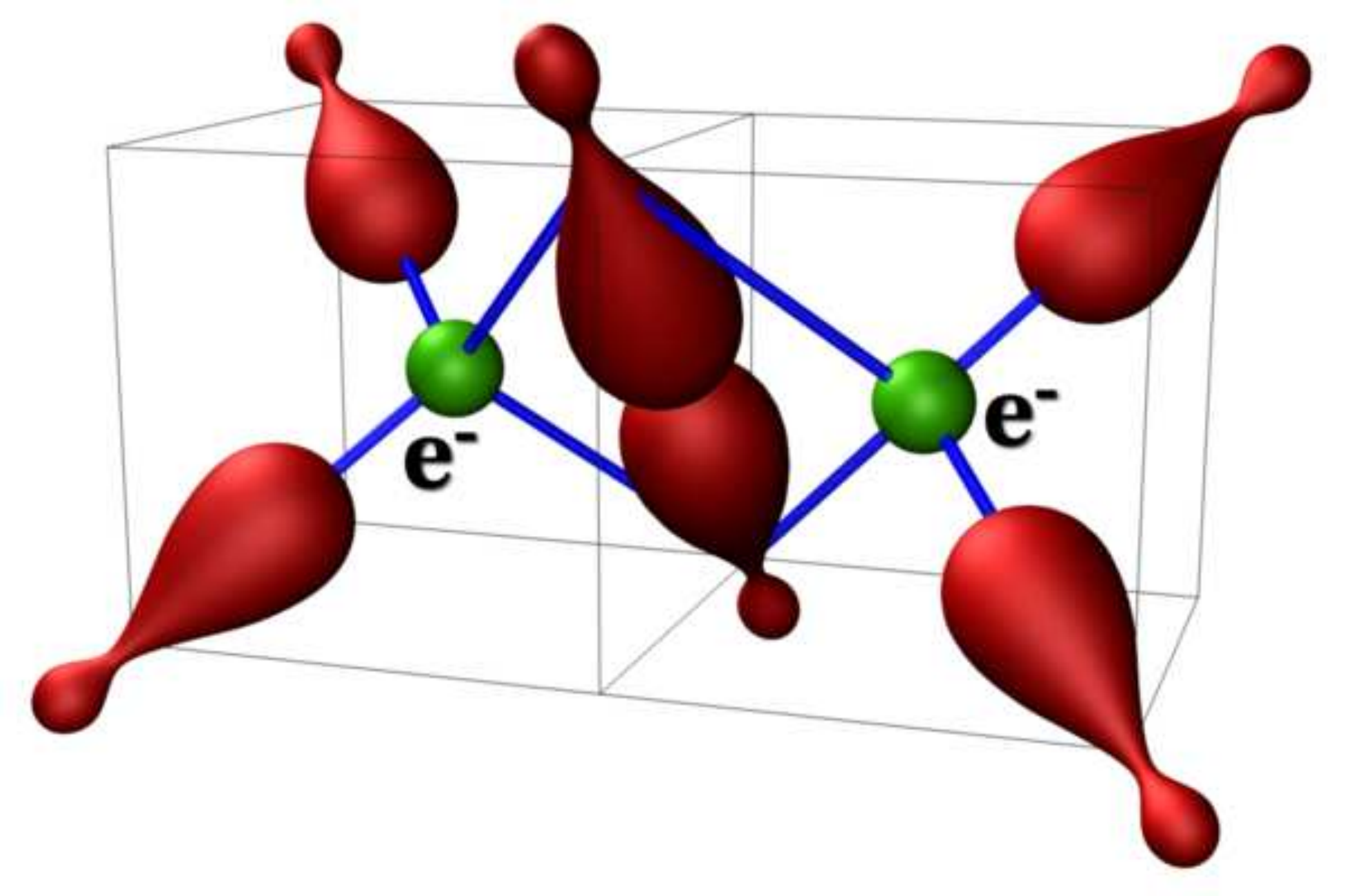}
\caption{Polarization clouds for a nearest neighbor electronic
  bipolaron. The two central ``shared'' As atoms have a polarization
  different from that of the usual polaron clouds. }
\label{fig10}
\end{figure}

A similar analysis applies for a nearest neighbor bipolaron. In this
case, as illustrated in Fig. \ref{fig10}, a total of 6 As atoms are
polarized. Four of them have clouds of the original type, while 2,
which are neighbors to both Fe atoms, have yet another type of
polarization cloud which lies in the plane perpendicular to be
bipolaron axis. Straightforward geometry shows that this polarization cloud
has magnitude $\sqrt{8\over 3}$ larger than an original cloud,
therefore one has to replace $g \rightarrow g\sqrt{8\over 3}$ (or
$4g^2 \rightarrow {32\over 3} 
g^2$) to
find the energy of each new cloud. 

\begin{figure}[t]
\includegraphics[width=0.5\columnwidth]{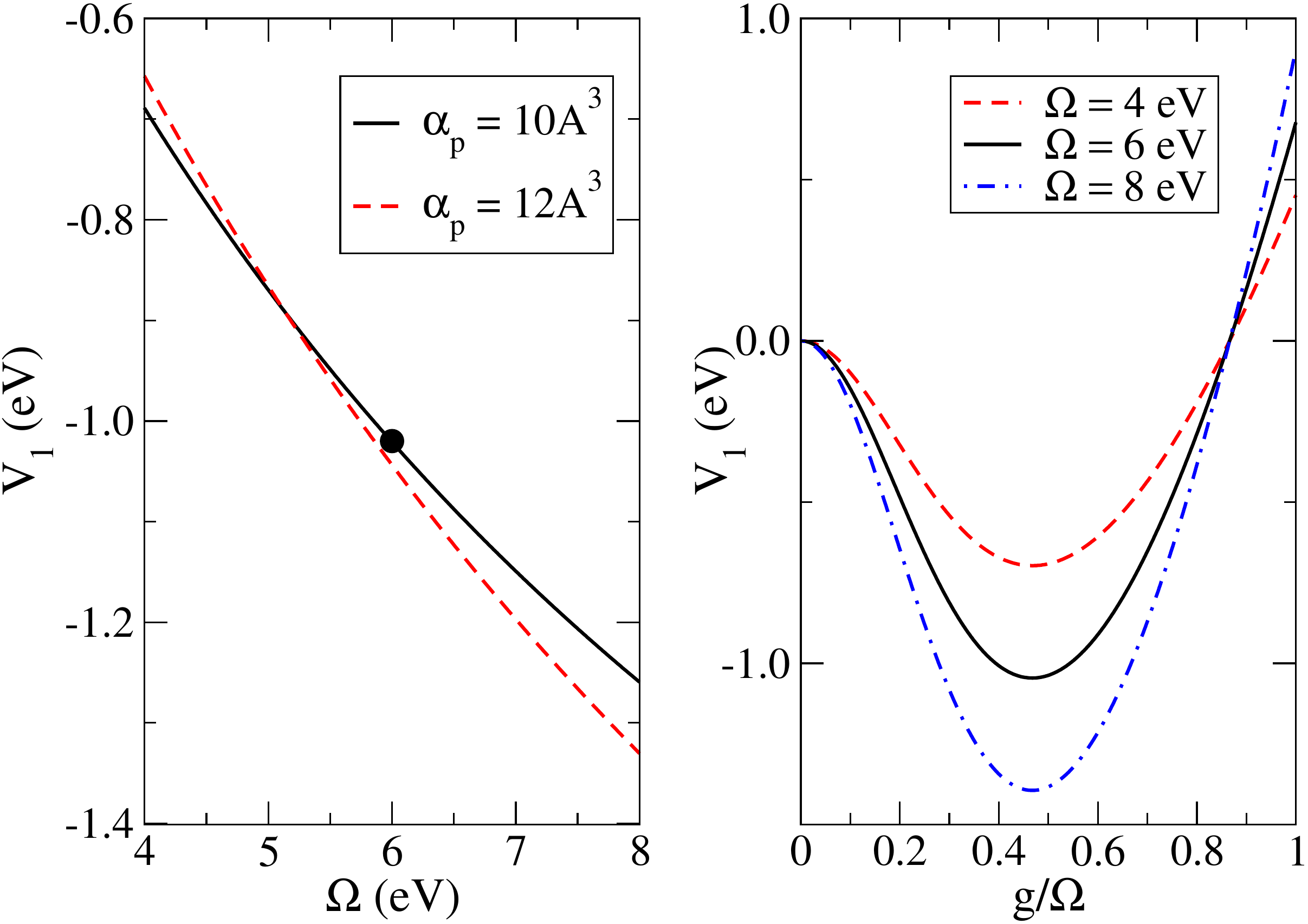}
\caption{Left panel: Effective nearest neighbor bipolaron interaction
  $U_1$ as a 
  function of $\Omega$ for two polarizability values. Right panel:
  $U_1$ vs. $g/\Omega$ for $\Omega = 4,6$ and 8 eV. }
\label{fig11}
\end{figure}

Thus, the total energy of a static nearest neighbor bipolaron is:
\begin{equation}
\EqLabel{be6}
E_{BP,1} = 4 \left[\Omega -\sqrt{\Omega^2+4g^2}\right] + 2\left[\Omega
  -\sqrt{\Omega^2+{32\over 3}g^2} \right]
\end{equation}
and the difference between this and the energy of two unbound polarons
is:
\begin{equation}
\EqLabel{b7}
U_1 = E_{BP,1}- E_{BP,\inf}= 4 \sqrt{\Omega^2+4g^2}-2\Omega - 2
\sqrt{\Omega^2+{32\over 3}g^2} 
\end{equation}
This effective energy is plotted in Fig. \ref{fig11} both as a function
of $\Omega$ for a fixed polarizability (left panel) and as a function
of $g/\Omega$ (right panel) for several values of $\Omega$.

The first interesting observation is that this effective interaction
is attractive, favoring a bound nearest neighbor bipolaron  over
two independent 
polarons. As shown in the right panel, this is only true for 
small $g/\Omega<1 $ values (if $g/\Omega$ becomes large enough, the
clouds saturate to their maximum allowed values and the two new clouds are
energetically less favorable than four original clouds). Remarkably,
the strongest binding energy is for $g/\Omega \approx 0.45$ for
reasonable values of $\Omega$, which is
close to the  values we expect for this parameter (see
Fig.~\ref{fig3}).  From this point of view, we can say that these
lattices are already very close to being fully optimized. As discussed for
the case of the 2nd 
nn bipolaron, it is possible to further change this interaction
by changing the lattice geometry: a smaller $\theta$ angle will again
increase this binding energy, but this is balanced by a
decrease  if the Fe-As distance is increased.

The creation operators for these new clouds are:
\begin{equation}
\EqLabel{b8}
\tilde{\tilde{\gamma}}^\dagger_{i,\pm,\sigma}=\cos \gamma s_{i,\sigma}^\dagger
- \sin \gamma 
	  {\pm p^\dagger_{i,\lambda,\sigma} + p^\dagger_{i,3,\sigma} \over
		\sqrt{2}}  
\end{equation}
where $\lambda=x$ or $y$ is the direction perpendicular to the
bipolaron axis, and the $\pm$  sign
shows whether the cloud is parallel/antiparallel to the $\lambda$ axis. The
new mixing angles are:
\begin{eqnarray}
\nonumber \cos \gamma &&= \sqrt{{1\over 2}\left(1+{\Omega\over
	\sqrt{\Omega^2+{32\over 3}g^2}}\right)}\\
\EqLabel{b9} \sin \gamma &&= \sqrt{{1\over 2}\left(1-{\Omega\over
	\sqrt{\Omega^2+{32\over 3}g^2}}\right)}
\end{eqnarray}

\begin{figure}[t]
\includegraphics[width=0.5\columnwidth]{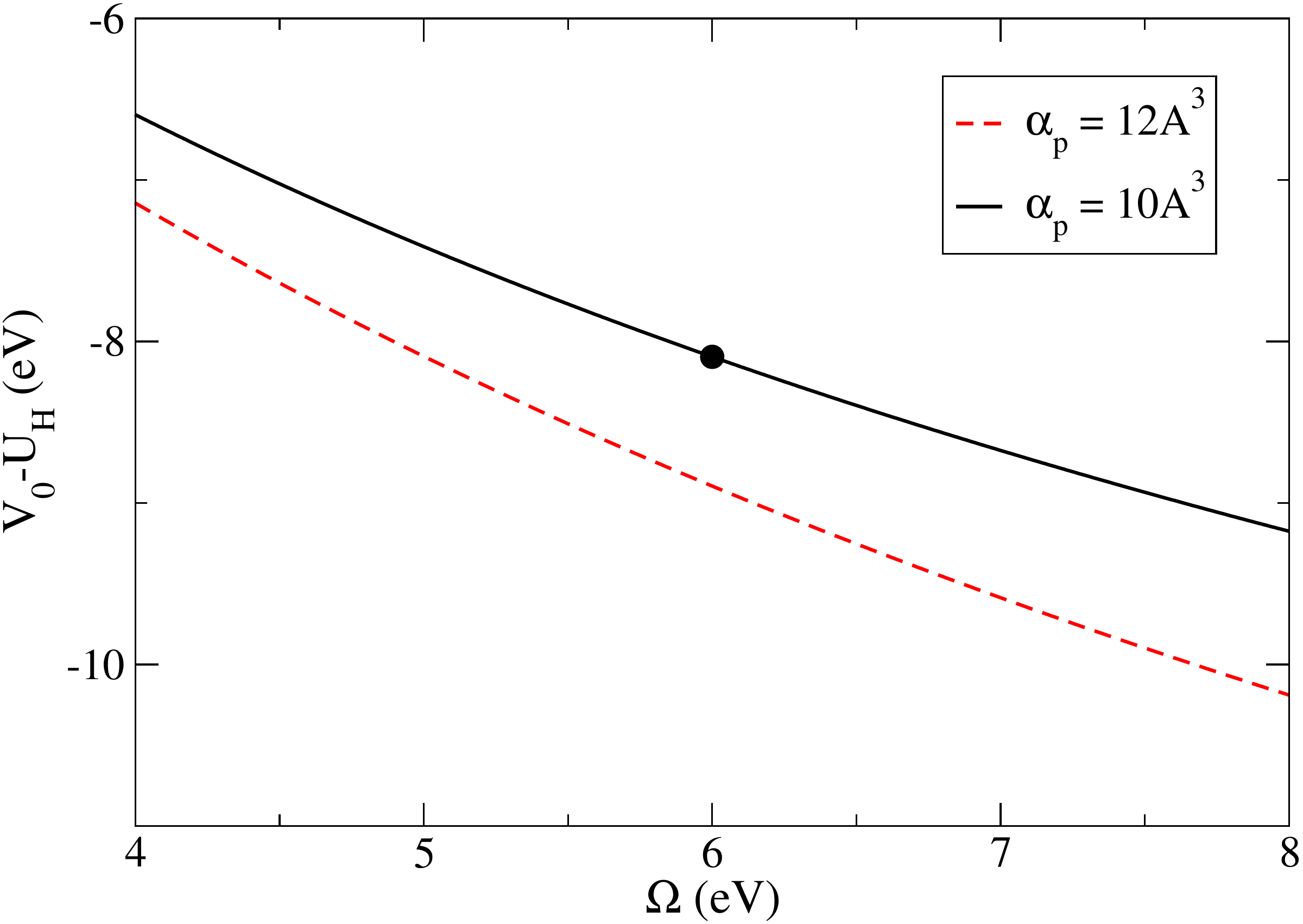}
\caption{Renormalization of the on-site interaction $U_0-U_H$
  vs. $\Omega$, for two values of the polarizability. This is always
a large  attractive contribution, which significantly screens the
  on-site Hubbard repulsion. The dot shows the typical value we use.}
\label{fig12}
\end{figure}

The analysis to this point would be identical if the static
bipolarons were in a triplet state. The last case, namely that of an onsite
bipolaron, is however only possible for a singlet state. In this
case, there are again only 4 polarization clouds, as for a
polaron. However, because the charge creating the electric field is
doubled, we now replace $g \rightarrow 2g$. As a result, the energy of
the on-site bipolaron is:
\begin{equation}
\EqLabel{b10}
E_{BP,0} = 4\left[\Omega - \sqrt{\Omega^2 +16g^2}\right]
\end{equation}
The effective on-site interaction is then:
\begin{equation}
\EqLabel{b11}
U_0 = U_{H} + E_{BP,0} - E_{BP,\infty} =  U_{\rm H}- 4\left[\sqrt{\Omega^2
	+16g^2} + \Omega - 2 \sqrt{\Omega^2 +4g^2}\right]
\end{equation}
where $U_H$ is the on-site Hubbard repulsion between the two
charges. Clearly, the polarization of neighboring As atoms always acts
to screen out the 
on-site repulsion, and this effect can be very significant. Indeed, in
Fig. \ref{fig12} we plot this renormalization energy, which is several
eV for typical parameters of interest to us.

One more comment is in order. In the linear regime, this
renormalization energy would be $\approx -16g^2/\Omega\approx -16eV$,
which is similar to the values used in Ref. \cite{George}. The smaller
values we find here are a direct consequence of the non-linear
effects, which are accounted for our this model.

The cloud operators for the on-site bipolaron are similar to the
original polaron cloud operators:
\begin{equation}
\EqLabel{b12}
\tilde{\tilde{\tilde{\gamma}}}^\dagger_{i,\lambda, \pm,\sigma}=\cos \delta
s_{i,\sigma}^\dagger 
- \sin \delta
	  \left[\pm \sin\theta p^\dagger_{i,\lambda,\sigma} + \cos \theta
		p^\dagger_{i,3,\sigma}\right] 
\end{equation}
where $\lambda=1$ or 2 depending on the in-plane alignment of the cloud,
and $\pm$ reflects its orientation. The only difference is the new
mixing angle:
\begin{eqnarray}
\nonumber \cos \delta&&= \sqrt{{1\over 2}\left(1+{\Omega\over
	\sqrt{\Omega^2+16g^2}}\right)}\\
\EqLabel{b13} \sin \delta &&= \sqrt{{1\over 2}\left(1-{\Omega\over
	\sqrt{\Omega^2+16 g^2}}\right)}
\end{eqnarray}

To summarize, this calculation reveals that for the parameters of
interest to us, we can have a stable on-site bipolaron if the Hubbard
repulsion is not too large. The nearest-neighbor bipolaron is
 bound, whereas the 2nd nearest neighbor bipolaron is unstable to
dissociation into two free polarons.

\subsection{First order perturbation: effective hopping integrals}

We continue the
discussion for singlet bipolarons, and analyze how the hopping Hamiltonian
mixes together the low energy bipolaron states $s^\dagger_{i,
  i+\vec{\delta}}|i, i+\vec{\delta}\rangle$. In this notation, $\vec{\delta}$
measures the distance between the two charges, and the state
$|i,i+\vec{\delta}\rangle$ describes the  As atoms when
there are charges at Fe sites $i$ and $i+\vec{\delta}$. Similar to the
states $|i\rangle$ defined for single polarons, the states
$|i,i+\vec{\delta}\rangle$  have all As atoms which are not nearest
neighbors to either Fe$_i$ or Fe$_{i+\vec{\delta}}$ in their ground
state, while the As atoms in the vicinity of the charged Fe atoms are
described by the appropriate $\gamma^\dagger,
\tilde{\gamma}^\dagger,\tilde{\tilde{\gamma}}^\dagger$ or
$\tilde{\tilde{\tilde{\gamma}}}^\dagger$ operator, as detailed in the
previous section. We do not list 
these expressions explicitly here, as they are rather long and tedious
but otherwise
straightforward.

From these states, we can define  a basis
of bipolaron states invariant to translations on the $N\times N$ unit
cells lattice:
\begin{equation}
\EqLabel{20}
|\vec{k}, \vec{\delta}\rangle = \sum_{i} {e^{i\vec{k}\cdot 
  (\vec{R}_i + {\vec{\delta}\over 2})}\over N}  s^\dagger_{i,
  i+\vec{\delta}}|i, i+\vec{\delta}\rangle. 
\end{equation}
Here $\vec{k}$ is the center-of-mass momentum of the bipolaron pair. It can
take the usual equally-spaced values inside the Brillouin zone allowed by the
periodic boundary conditions. However, some care needs to be taken in
the definition of $\vec{\delta}$, since if we take all its possible
values, we double-count the states (for example,
$|\vec{k},\vec{e}_x\rangle = |\vec{k},-\vec{e}_x\rangle$, etc).  This
is due to the fact that $\vec{\delta}$ is like a nematic vector, with
a size and orientation but
without a pointing arrow. 
Let $\vec{\delta}=
\delta_x \vec{e}_x + \delta_y \vec{e}_y$, and for simplicity assume
that the lattice size is an odd number, $N=2n+1$. Then, double-counting is
avoided provided that when $\delta_x = 0$, we allow
$\delta_y = 0,1, \dots ,n$ (because of the periodic boundary conditions,
the  distance between charges cannot be more than half the
dimension of the system). If $\delta_x=1,2,\dots,n$, then $\delta_y = 
-n , ..., n$. Both positive and negative values are needed in this
case, because, for example,  the $|\vec{k}, \vec{e}_x+\vec{e}_y\rangle$
state is distinct 
from the $|\vec{k}, \vec{e}_x-\vec{e}_y\rangle$ state, however these
are the only two 
distinct states of the 2nd nearest neighbor bipolarons. In conclusion, for a
$N\times N$ lattice with $N=2n+1$, there are $(n+1)(2n+1)-n$ distinct
singlet eigenstates corresponding to a given  total momentum $\vec{k}$
(the remaining are triplet states. Similar accounting can be done for
lattices with even number of sites).

Because our total Hamiltonian is also invariant to translations, it
will not mix states with different momenta. Therefore, we can solve
the problem separately in each $\vec{k}$ subspace, and only need to
calculate the various $\langle 
\vec{k}, \vec{\delta}'| {\cal H} | \vec{k}, \vec{\delta}\rangle$
matrix elements. There are three sets of such contributions. First,
there are ``diagonal''
contributions, already discussed in the previous
section. Specifically, $\langle 
\vec{k}, \vec{\delta}| {\cal H} | \vec{k}, \vec{\delta}\rangle = \langle 
\vec{k}, \vec{\delta}| \hat{U}+ {\cal H}_{\rm As} + {\cal H}_{\rm int} |
\vec{k}, \vec{\delta}\rangle  = U_{\vec{\delta}}$ where $U_0$, $U_1$
and $U_2$ of Eqs. (\ref{b4}), (\ref{b7})  and (\ref{b11})
correspond to $\vec{\delta}$ indicating, respectively, an on-site, a
nearest neighbor or a 2nd nearest neighbor bipolaron state. If
$|\vec{\delta}|\ge 2$, then $U_{\vec{\delta}}=0$, since we measure
the interactions energies with respect to the energy of two unbound static
polarons. 

The nn hopping Hamiltonian links
states with the same momentum but bipolaron distances
$\vec{\delta}$ and $\vec{\delta}'$ which are nearest neighbor, {\em
  i.e.} $|\vec{\delta}-\vec{\delta}'|=1$. We therefore also need to
calculate these matrix elements $\langle 
\vec{k}, \vec{\delta}\pm \vec{e}_{x/y}| {\cal H} | \vec{k},
\vec{\delta}\rangle= \langle 
\vec{k}, \vec{\delta}\pm \vec{e}_{x/y}| \hat{T} | \vec{k},
\vec{\delta}\rangle$.  Similarly, for finite $t'$ there are matrix elements of
$\hat{T}'$ between 2nd nearest neighbor states with
$|\vec{\delta}-\vec{\delta}'|=\sqrt{2}$. Taken together, these three
sets  exhaust all
matrix elements which 
are finite and therefore we can 
diagonalize the resulting matrix and find the energies and
eigenstates corresponding to any desired $\vec{k}$, within first order
perturbation theory.  

If $\vec{\delta}$ is large enough, the hopping Hamiltonian $\hat{T}$
will hop one of the electrons away from its clouds, while the second one
remains unperturbed. As a result, one expects the matrix element to be
proportional to the $t_{\rm eff}$ of the single polaron. Indeed,
explicit calculation shows that if the separation $\vec{\delta}$ is at
least that of a  3rd nearest neighbor bipolaron, then:
\begin{eqnarray}
\nonumber
 &&\langle
\vec{k}, \vec{\delta}+\vec{e}_x| \hat{T} | \vec{k}, \vec{\delta}\rangle= - 2
t_{\rm eff} \cos {k_x a\over 2}\\
 &&\EqLabel{bb1}
\langle \vec{k}, \vec{\delta}\pm\vec{e}_y| \hat{T} | \vec{k},
 \vec{\delta}\rangle= - 2 
t_{\rm eff} \cos {k_y a\over 2}
\end{eqnarray}
The cosine factors appear because of the phases in the definitions of the
 eigenstates $| \vec{k}, \vec{\delta}\rangle$, which change
 when the singlet separation $\vec{\delta}$ changes. Given the
 restrictions on the allowed values of $\vec{\delta}$, one needs to be
 a bit careful near the boundary of allowed values. For example, based
 on the definition of Eq. (\ref{bb1}), it is straightforward to check
 that 
$$
| \vec{k}, \delta_x \vec{e}_x + (n+1)\vec{e}_y\rangle = e^{i
  k_y a{2n+1\over 2}}|\vec{k},  \delta_x \vec{e}_x -n
\vec{e}_y\rangle.
$$
Given the periodic boundary conditions, the allowed values for
momentum, if $N=2n+1$, are $k_y a= 2\pi n_y/(2n+1)$, and 
the phase factor linking the two states can only be $\pm1$. This sign is
important because hopping in the $y$-direction links the state with
$\delta_y =n$ 
to that with $\delta_y =n+1$, however the later one is not
amongst the allowed $\vec{\delta}$ values. The identity above links it
to
an allowed value, but also introduces a phase shift (the sign) which
is transferred 
to the corresponding matrix element. A similar situation arises for
hopping in the $x$ direction when $\delta_{x} = n$, $\delta_x=0$ and
hopping in the $y$ direction for 
$\delta_y=-n$. Finally, it is important to 
note that these phase shifts are different for triplet states (they
have opposite sign in most cases, but not all). Other than this (and
the non-existence of an triplet onsite 
bipolaron, as already discussed), essentially everything else is the
same for triplet states. 
%

Different effective hopping integrals appear when the electrons are
closer together. For nearest-neighbor hopping, there are four such
special cases. The first corresponds to 
hopping from 3rd nearest neighbor to nearest neighbor bipolaron. A
straightforward calculation of the overlap of the corresponding
polarization clouds leads to an effective hopping:
\begin{eqnarray}
\nonumber t_3 = &&t\langle i, i+\vec{e}_x| i, i+2\vec{e}_x\rangle = t \left[\cos \alpha(\cos^2 \alpha + \sin^2 \alpha \sin^2
  \theta) \right]^4\\
 &&\EqLabel{bb2}
\times\left[\cos \alpha \cos \gamma + \sin \alpha \sin \gamma
  \left({\cos \theta\over \sqrt{2} } +{\sin \theta\over 2}\right)\right]^4
\end{eqnarray}
The matrix elements for such processes are:
\begin{eqnarray}
\nonumber
 &&\langle
\vec{k}, 2\vec{e}_x| \hat{T} | \vec{k},\vec{e}_x \rangle= - 2
t_3 \cos {k_x a\over 2}\\
 &&\EqLabel{bb3}
\langle \vec{k}, 2\vec{e}_y| \hat{T} | \vec{k},
\vec{e}_y\rangle= - 2 
t_{3} \cos {k_y a\over 2},
\end{eqnarray}
in other words like the general matrix elements of
Eq. (\ref{bb1}), except with the appropriate value of the effective
hopping integral.

The second special case involves  hopping 
between 2nd and 4th nn bipolarons. In this case, the
overlap between the corresponding polarization clouds leads to: 
\begin{eqnarray}
\nonumber t_4 = && t \langle i, i+\vec{e}_x+\vec{e}_y|i,
 i+\vec{e}_x+2\vec{e}_y\rangle = t \cos^6 \alpha \left[\cos^2 \alpha +
 \sin^2 \alpha \cos^2
  \theta\right]^4 \\
 &&\EqLabel{bbb4}\times\left[\cos \alpha\cos\beta + \sin
 \alpha\sin\beta\cos\theta\right]^2 
\end{eqnarray}
and the matrix elements:
\begin{eqnarray}
\nonumber
 &&\langle
\vec{k}, 2\vec{e}_x\pm\vec{e}_y| \hat{T} | \vec{k},\vec{e}_x\pm
 \vec{e}_y \rangle= - 2 
t_4 \cos {k_x a\over 2}\\
 &&\EqLabel{bbb5}
\langle \vec{k}, \vec{e}_x\pm 2\vec{e}_y| \hat{T} | \vec{k},
\vec{e}_x\pm\vec{e}_y\rangle= - 2 
t_{4} \cos {k_y a\over 2}.
\end{eqnarray}

The third special case involves  hopping 
between 2nd and 1st nearest neighbor bipolarons. In this case, the
overlap between the polarization clouds corresponding to these two
cases leads to a renormalized hopping: 
\begin{eqnarray}
\nonumber t_2 = && t \langle i, i+\vec{e}_x|i,
 i+\vec{e}_x+\vec{e}_y\rangle = t \cos^6 \alpha \left[\cos^2 \alpha +
 \sin^2 \alpha \cos^2 
  \theta\right]^2 \left[\cos \beta \cos \gamma + {\sin \beta \sin
	\gamma\over \sqrt{2}}\right]^2\\
 &&\EqLabel{bb4}\times \left[\cos \alpha \cos \gamma + \sin \alpha \sin \gamma
  \left({\cos \theta\over \sqrt{2} } +{\sin \theta\over 2}\right)\right]^2
\end{eqnarray}
and the matrix elements:
\begin{eqnarray}
\nonumber
 &&\langle
\vec{k}, \vec{e}_x\pm \vec{e}_y| \hat{T} | \vec{k},\vec{e}_y \rangle= - 2
t_2 \cos {k_x a\over 2}\\
 &&\EqLabel{bb5}
\langle \vec{k}, \vec{e}_x\pm \vec{e}_y| \hat{T} | \vec{k},
\vec{e}_x\rangle= - 2 
t_{2} \cos {k_y a\over 2}.
\end{eqnarray}

Finally, for the hopping between on-site and nearest neighbor
bipolaron, the effective hopping is
found to be: 
\begin{equation}
\EqLabel{bb6}
t_1 =t\langle i,i| i, i+\vec{e}_x\rangle =  t \cos^4 \alpha
\cos^4(\alpha-\delta) \left[\cos \delta \cos \gamma + \sin \delta \sin
  \gamma 
  \left({\cos \theta\over \sqrt{2} } +{\sin \theta\over 2}\right)\right]^4
\end{equation}
and the matrix elements are:
\begin{eqnarray}
\nonumber
 &&\langle
\vec{k}, \vec{e}_x| \hat{T} | \vec{k}, 0\rangle= - 2\sqrt{2}
t_1 \cos {k_x a\over 2}\\
 &&\EqLabel{bb7}
\langle \vec{k}, \vec{e}_y| \hat{T} | \vec{k}, 0\rangle= - 2 \sqrt{2}
t_{1} \cos {k_y a\over 2}.
\end{eqnarray}
The extra factor of $\sqrt{2}$ is because of the on-site singlet
normalization, see Eq. (\ref{b1}). Of course, for triplet states
$t_1=0$, since there is no on-site triplet. 

\begin{figure}[t]
\includegraphics[width=0.5\columnwidth]{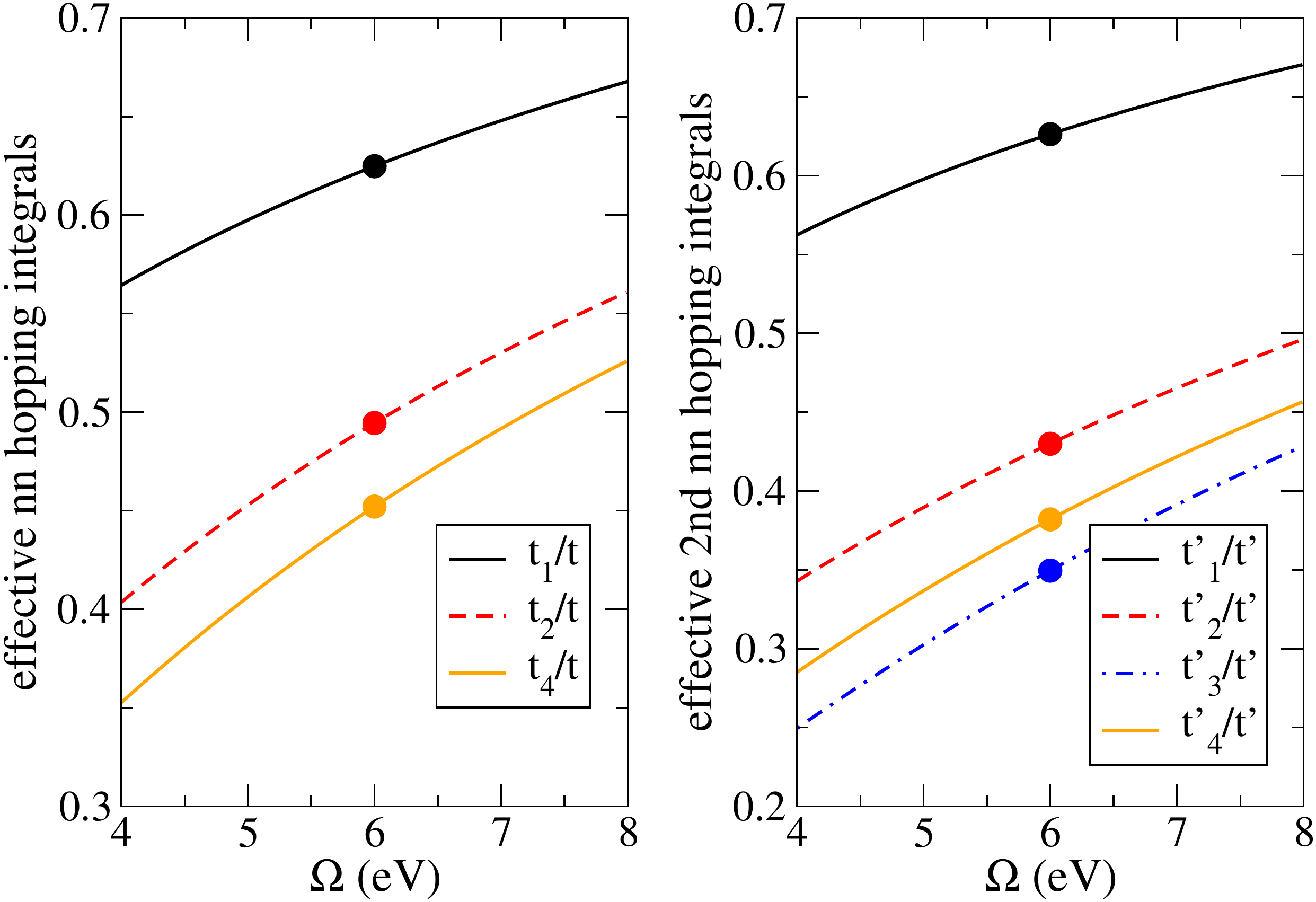}
\caption{Effective hoppings for closely-spaced bipolarons. For these
  parameters, the $t_2$ and $t_3$ curves are essentially superimposed,
  therefore we only show $t_2/t$. }
\label{fig12b}
\end{figure}

In Fig. \ref{fig12b}(a) we plot the values of $t_1/t, t_2/t, t_3/t$
and $ t_4/t$ 
vs. $\Omega$ for $\alpha_P=10\AA^3$. These should be compared with
$t_{\rm eff}/t$ shown in Fig. \ref{fig6}. The overall changes are
relatively small.

The situation for second nearest neighbor hopping is similar. If
$\vec{\delta}$ is large enough, then the cloud overlap is the same as
for individual polarons and the renormalized hopping is $t'_{\rm
  eff}$, giving the matrix elements:
\begin{equation}
\EqLabel{bb8}
\langle
\vec{k}, \vec{\delta}+\vec{e}_x\pm\vec{e}_y| \hat{T}' | \vec{k},
\vec{\delta}\rangle= - 2 
t'_{\rm eff} \cos {(k_x\pm k_y) a\over 2}
\end{equation}
Again, special care must be used for values of $\vec{\delta}$ near the
edges of its area of allowed values, so that the periodic boundary
conditions are properly accounted for. This is done in a manner
totally analogous to that explained for nearest neighbor hopping.

There are again four special cases of different renormalization
of $t'$. One involves the hopping between the two possible
nearest-neighbor bipolaron states, which has the matrix element:
\begin{equation}
\EqLabel{bb9}
\langle
\vec{k}, \vec{e}_x| \hat{T}' | \vec{k},
\vec{e}_y\rangle= - 2 
t'_2\left[ \cos {(k_x+ k_y) a\over 2}+\cos {(k_x- k_y) a\over 2}\right]
\end{equation}
where
\begin{equation}
\EqLabel{bbe9}
t'_2 = t'\langle i, i+x|i, i+y\rangle= t'\left[\cos
  \alpha\right]^8\left[ \cos\alpha \cos\gamma +\sin \alpha \sin \gamma
  \left({\cos\theta\over \sqrt{2}} + {\sin \theta\over
	2}\right)\right]^4 \left[\cos^2\gamma + {1\over 2}
  \sin^2\gamma\right]^2 
\end{equation}
The second involves the hopping between $2^{nd}$ and $3^{rd}$ nn
bipolarons, as well as hopping away from a $2^{nd}$ nn bipolaron,
with the matrix elements: 
\begin{eqnarray}
\nonumber
 &&\langle
\vec{k}, 2\vec{e}_x\pm 2\vec{e}_y| \hat{T}' | \vec{k},
\vec{e}_x\pm \vec{e}_y\rangle= - 2 
t'_3\cos {(k_x\pm k_y) a\over 2}\\
 &&\nonumber
\langle\vec{k}, 2\vec{e}_x| \hat{T}' | \vec{k},
\vec{e}_x\pm \vec{e}_y\rangle= - 2 
t'_3\cos {(k_x\mp k_y) a\over 2}\\
 &&\EqLabel{bba9}
\langle\vec{k}, 2\vec{e}_y| \hat{T}' | \vec{k},
\vec{e}_x\pm \vec{e}_y\rangle= - 2 
t'_3\cos {(k_x\mp k_y) a\over 2}\\
\end{eqnarray}
where $t'_3 = t'\langle i, i+x+y|i, i+2x+2y\rangle=t'\langle
i+x+y|i+2x\rangle$ is found to be:
\begin{equation}
\EqLabel{bba9}
t'_3= t'\left[\cos
  \alpha\right]^{10}\left[ \cos\alpha \cos\beta+\sin \alpha \sin
  \beta\cos\theta 
  \right]^2 \left[\cos^2\alpha++\sin^2\alpha \cos(2\theta)\right]^2 
\end{equation}
The third case involves hopping between $1^{st}$ and $4^{th}$ nn
bipolarons, with matrix elements:
\begin{eqnarray}
\nonumber
 &&\langle
\vec{k}, 2\vec{e}_x\pm \vec{e}_y| \hat{T}' | \vec{k},
\vec{e}_x\rangle= - 2 
t'_4\cos {(k_x\pm k_y) a\over 2}\\
 &&\EqLabel{bbc9}
\langle\vec{k}, \vec{e}_x\pm2\vec{e}_y| \hat{T}' | \vec{k},
\vec{e}_y\rangle= - 2 
t'_4\cos {(k_x\pm k_y) a\over 2}\\
\end{eqnarray}
where $t'_4 = t'\langle i+x|i+2x+y\rangle$ equals:
\begin{equation}
\EqLabel{bbd9}
t'_4= t'\left[\cos
  \alpha\right]^{8}\left[\cos^2\alpha +\sin^2 \alpha
  \cos(2\theta)\right]^2\left[\cos\alpha\cos\gamma
  +\sin\alpha\sin\gamma\left({\cos\theta\over\sqrt{2}} +
  {\sin\theta\over 2}\right)\right]^4.
\end{equation}

Finally, the hopping between  an on-site and a 2nd nearest
neighbor bipolaron (again, possible only for singlet states) leads to:
\begin{equation}
\EqLabel{bbe8}
\langle
\vec{k}, \vec{e}_x\pm\vec{e}_y| \hat{T}' | \vec{k},
0\rangle= - 2 \sqrt{2}
t'_1 \cos {(k_x\pm k_y) a\over 2}
\end{equation}
where
\begin{equation}
\EqLabel{bb10}
t'_1 = t'\langle i, i|i, i+x+y\rangle= t'\left[\cos\alpha
  \cos(\alpha-\delta)\right]^6 
\left[\cos(\beta-\delta)\right]^2.
\end{equation}
The renormalized values $t'_1, t'_2, t'_3$ and $t'_4$ are shown, in
units of $t'$, in 
Fig. \ref{fig12b}(b).

The ensemble of all these matrix elements define the matrix that needs
to be diagonalized in order to find eigenstates and eigenvalues for a
given total momentum $\vec{k}$. The results are shown in the main
paper.

\section{Estimates for the accuracy of various approximations}

In the previous sections we argued that first order perturbation
theory is reasonably accurate for our Hamiltonian, given the values of the
various parameters relevant for the Fe-based superconductors. However,
the Hamiltonian itself already embodies two approximations, namely (i)
that only As atoms nn to a doping charge become polarized, and (ii)
that dipole-dipole interactions between the various polarized As atoms
are ignored.

We provide here rough estimates for the accuracy of these
approximations. We begin by discussing the
single polaron.

Assume that interactions with the 8 As atoms which are 2$^{nd}$ nn
to a charge are also included. This supplementary interaction would be
described by a Hamiltonian similar to ${\cal H}_{\rm int}$ of
Eq. (\ref{1}), except it would have a new energy scale $g'$ and new
angles describing the orientations of these Fe-As bonds.

From the definition of the interaction energy, Eq. (\ref{e6}), we
have:
$$
{g'\over g} = {R^2\over R'^2}
$$
where $R'= \sqrt{11}a/2$ is the distance from an Fe to a 2$^{nd}$ nn
As atom. This comes about because the only difference is in the
electric fields created at the two sites, and these are inversely
proportional to the square of the distances.

It then follows immediately that the contribution to the static
polaron energy of these 8 clouds would be:
\begin{equation}
\EqLabel{n1}
E_{\rm corr} = 8 \left( \Omega - \sqrt{\Omega^2 + 4 g'^2}\right),
\end{equation}
which for our typical parameters is $E_{\rm corr} =-1.2$ eV. Of
course, considering the polarization of even more distant As atoms
would lower this energy even more.

However, this decrease is more than compensated for by consideration of
dipole-dipole interactions. To estimate these energies, let us
consider only the 6 pair interactions between the 4 largest dipole moments
on the nn As sites. Consider a pair of two such dipoles, with moments
$\vec{p}_1$ and $\vec{p}_2$.
Their magnitude is given by Eq. (\ref{e11}) and
their orientations are 
straightforward to determine, as is the distance $\vec{d}$ between them.
Using these values, we find for any such pair
that:
\begin{equation}
\EqLabel{n2}
E_{\rm d-d} = {1\over d^3} \left(\vec{p}_1\cdot \vec{p}_2 -
3{(\vec{p}_1\cdot \vec{d})(\vec{p}_2\cdot \vec{d})\over
  d^2}\right)={10 \alpha_p \Omega\over 3\sqrt{2} a^3} {g^2\over
  \Omega^2+4g^2} 
\end{equation}
where the numerical factor in front is due to geometrical
considerations. For our typical parameters, $E_{\rm d-d} =0.66$eV and
therefore the total for the 6 pair interactions is $\approx 4$ eV. Inclusion of
contributions from interactions with the dipoles on the  2$^{nd}$ nn
As atoms will
decrease this, because nn and 2$^{nd}$ nn dipoles closest to each
other  are now
roughly parallel, not anti-parallel, to each other. 

Taken together and without further corrections, these two energies
would renormalize $E_{B,PS}$ to about half the value predicted in
their absence. If we keep adding further rows of neighbors
into the calculation, the polaron energy saturates (slowly) to around
-1.6eV (of course, if the fields decrease as $1/R^2$, the energy
slowly become more negative and will
eventually go to $-\infty$. However one has to take into
account that there are other screening mechanism that further screen
the interaction).  Having extended clouds would also lower the
effective hoppings somewhat more, although we expect this to be a much
smaller effect: clouds which are far from the electronic charge are
changed much if the charge moves by one site. As a result, their
 contribution to the total overlap should be very close to 1. 

Although the renormalization of the static part of the polaron energy is
quite substantial, these corrections do not have nearly
as large an  
effect on the bipolaron interaction energies $U_0, U_1$ and $U_2$. The
reason is that these are differences between the true bipolaron
energy and the energy of two free polarons, and therefore  many
contributions from both the extended clouds and the 
dipole-dipole interactions cancel each other out.

For example, consider the effects of these corrections on $U_1$, which
is the most interesting energy. For two polarons at an infinite
distance from each other, the dipole-dipole correction is 12 $E_{\rm
d-d}$ and there are a  total of 16 $2^{nd}$ nn extra clouds.  For a nn
bipolaron, the dipole-dipole energy is roughly 11 $E_{\rm d-d}$,
because the two As tetrahedra share one side (of course, the two
central As atoms have somewhat different dipole moments pointing in
somewhat different directions, and also there are smaller
contributions from interactions between the non-central As atoms
belonging to different Fe, but 11 $E_{\rm d-d}$ should be a reasonable
guess). This configuration would only have 12 $2^{nd}$ nn extra
clouds. Subtracting these corrections leads to an overall correction
of $U_1$ roughly equal to $-{1\over 2}E_{\rm corr} -E_{\rm d-d}
\approx 0$. In reality, the energy is lowered because the close
location of the two charges implies larger electric fields at all
sites for the nn bipolaron, therefore larger clouds. We find that as
we increase the number of 
As neighbors included in the cloud, $U_1$ is lowered from $\approx -1$
eV (the value we find when the approximations are made) to $-2.87$ eV
if we include 2nd 
nn and dipole-dipole interactions, to $-4eV $ if more
and more neighbors are added. Again, one expects that the electric fields will
be screened over some finite distance which will decide the precise
value. The point, though, is that relaxing the approximations increases
this attraction energy, in other words the physics we discuss here
becomes more relevant, not less.

\section{Link between BCS-like theory and experimental data}

There have been many works in the recent experimental literature
pointing to correlations between the critical temperature T$_c$ and
shorter Fe-As distances and smaller Fe-As-Fe angles, see for example
Ref. \onlinecite{R3b}. Such correlations would come as a natural
consequence in our theory, where the ``glue'' leading to  pair
formation is the polarization of nearby As ions, and thus it is very
sensitive to the precise location of the As with respect to Fe. 

As discussed towards the end of our main paper, we currently envision
two possible scenarios explaining the high-T$_c$ in the materials,
namely a BCS-like and a BEC-like scenario.  The BEC-like scenario
assumes that the nearest-neighbor attraction $U_1$ is larger than the
nearest-neighbor bare Coulomb repulsion (not included in our
simulations) and that pre-formed bosonic pairs appear well above
T$_c$. As discussed, there is experimental 
evidence consistent with this scenario. The BCS-like scenario assumes
that the nn Coulomb 
repulsion is large enough to overcome $U_1$, and therefore there are
no pre-formed pairs in the system. However, like in regular
superconductors, the existence of an attraction mechanism (given by
$U_1$), when combined with the existence of a Fermi sea and proper
inclusion of retardation effects, should suffice to lead to
Cooper-pair formation at T$_c$. 

Currently, our calculations are not detailed enough to allow us to
distinguish which scenario is the more likely one. In this section we
first {\em speculate} on the possible fit of the experimental data to
the BCS-like scenario, and then comment on the relevance of the same
data for the BEC scenario.

The well-known BCS formula links T$_c$ to the typical scale of the
attractive potential $V$ in an exponential dependence, $T_c \sim
\exp\left(-{1\over |V| \rho}\right)$, where $\rho$ is the density of
states at the Fermi energy. On a logarithmic scale, this implies $\ln
T_c = const. -{1\over |V|\rho}$, if we assume that the prefactor is
roughly constant. For simplicity, let us also assume that $V\propto
U_1 \propto \alpha_p \vec{E}_1 \cdot \vec{E}_2\propto \alpha_p
{\cos{\gamma}\over R^4}$, where $R$ is the Fe-As distance and $\gamma$
is the nn Fe-As-Fe angle. As explained in the main paper, this
expression gives the linear approximation for the nn $U_1$ attraction
~\cite{George}, and we use it here for simplicity -- since we will
ignore changes in the density of states for the various materials,
taking into consideration the small non-linear effects is superfluous.

\begin{figure}[b]
\includegraphics[width=0.60\columnwidth]{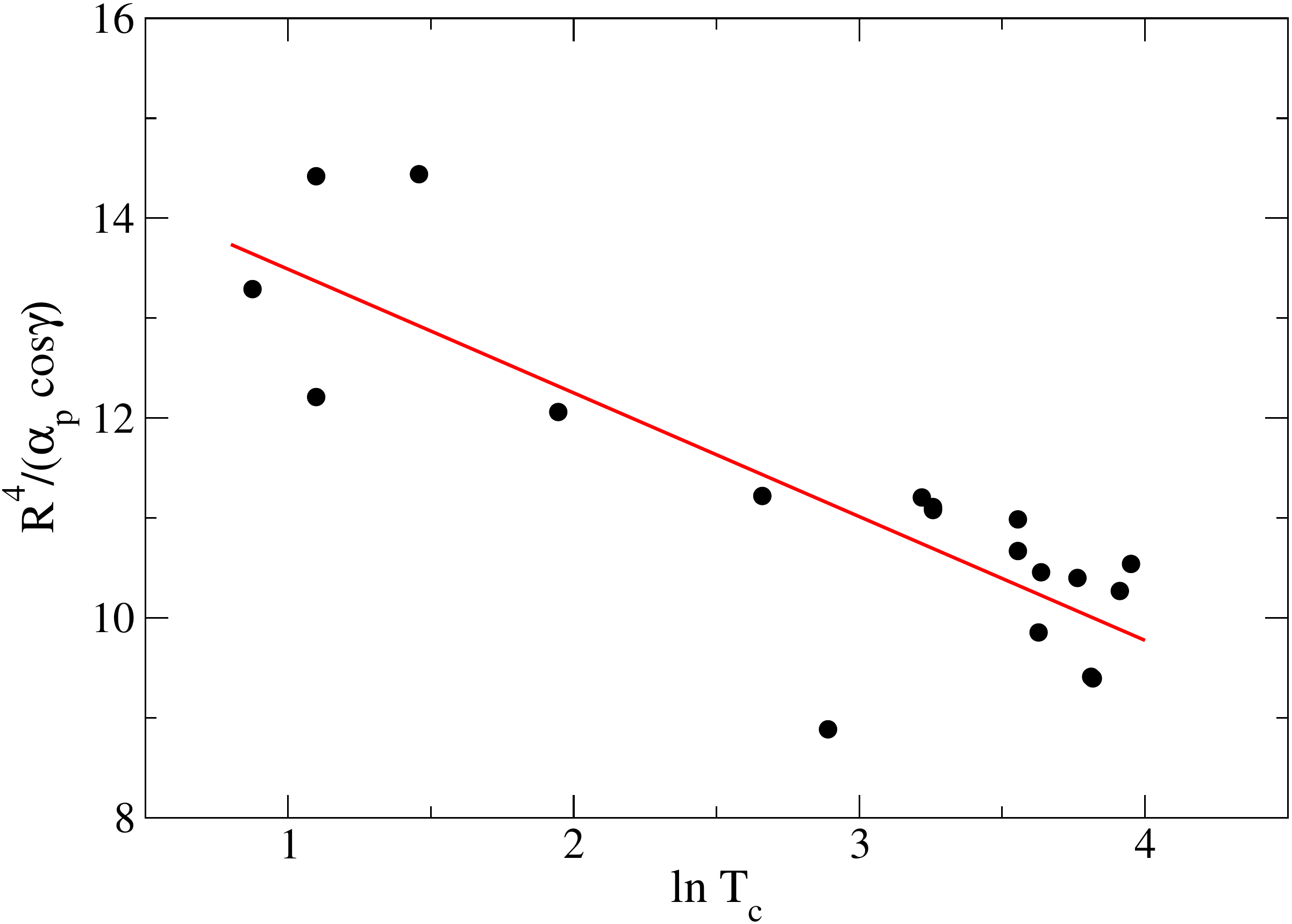}
\caption{Linear fit of $\ln T_c$ vs. $R^4\over \alpha_p
{\cos{\gamma}}$. The data points are taken from
Refs. \onlinecite{R3b,R1,R2,R3,R4,R5,R6,R7,R8,R9,R10,R11,R12,R13}.  }  
\label{fig13}
\end{figure}

In Fig. \ref{fig13} we show a plot of
T$_c$ vs. $R^4\over \alpha_p
{\cos{\gamma}}$ for a multitude of materials, which have
polarizabilities, $R$ and $\gamma$ values varying over fairly large
ranges~\cite{R3b,R1,R2,R3,R4,R5,R6,R7,R8,R9,R10,R11,R12,R13}. While
there is some spread to the data, it is not inconsistent with a linear
dependence like that predicted by a BCS formula. Of course, a lot more
theoretical work needs to be done before one could justify this
plot as more than just an accident.

This sensitivity of T$_c$ to $\alpha_p, R$ and $\gamma$ is not
inconsistent with a BEC-like scenario, either. Assuming that $U_1$ is
large enough that 
pre-formed pairs exist up to very high temperatures, then T$_c$ would
be related to the on-set of coherence between these bosons. Generally
speaking, this is achieved when the de Broglie wavelength becomes
comparable to the average spacing between bosons. Whatever the
specific relationship is, it will involve the effective mass of these
pre-formed pairs. This effective mass depends in a non-trivial fashion
on $\alpha_p, R$ and $\gamma$, as discussed in the main paper. This
dependence is not only through $U_1$, but to a large degree also
through the cloud overlaps determining the various $t_{\rm eff}$
values. Since all these quantities are very sensitive to $\alpha_p, R$
and $\gamma$, this scenario would also be compatible with a T$_c$
which depends on these quantities. Trying to fit it would, at this
point, be even more speculative than the fit shown in
Fig. \ref{fig13}. 

In any event, the underlying idea is that if the As
ions played a small, secondary-type role in determining the behavior
of these materials, this dependence would be very puzzling. In our
model, the As anions are the main enablers of the pairing mechanics,
therefore this sensitivity of T$_c$ to specific details of the lattice
structure follows quite naturally.